\documentclass{aa}

\usepackage{graphicx}
\usepackage{txfonts}
\usepackage{array,booktabs,tabularx}
\usepackage{ragged2e}
\usepackage{placeins}

\begin{document}

   \title{Beating and coupling in pulsating stars: A unified Fourier description and observational diagnostics
}
   \titlerunning{Beating and coupling in pulsating stars}

   \author{J\'ozsef M. Benk\H{o}
   and
   Emese Plachy
   }

   \institute{Konkoly Observatory, HUN-REN Research Centre for Astronomy and Earth Sciences,
              MTA Centre of Excellence, \\ Konkoly Thege u. 15-17., 1121 Budapest, Hungary, 
              \email{benko@konkoly.hu}
             }

   \date{Received Aug 20, 2026; accepted Sep 25, 2026}

\abstract
   {Long-term amplitude and phase variations in pulsating stars produce side peaks or multiplets in the Fourier spectrum of the light curves around the pulsation frequency and its harmonics. Such structures are often attributed to modulation, although linear beating and the non-linear coupling of close oscillations can also produce similar patterns.}
   {We developed a unified analytical Fourier description of these alternatives and identify observational relations that retain discriminatory power when their frequency patterns overlap.}
   {We represented the signal as a dominant, non-sinusoidal oscillation plus one or more secondary oscillations close to it, and analytically derived the frequencies, complex amplitudes, and phases resulting from linear superposition, quadratic coupling, and general non-linear coupling. Synthetic light curves illustrate these results. }
   {We implemented \textsc{freqdiag}, which compares modulation with primary--secondary quadratic coupling models anchored to either the higher- or lower-frequency side peak, using the same normalised complex amplitudes, and supplements the fit with hierarchical tests of side-peak amplitude progression. A sinusoidal secondary signal produces only one additional peak in the linear-beating case, whereas a non-sinusoidal secondary signal produces a harmonic sequence whose separation from the primary harmonics increases with order. Quadratic coupling produces terms $nf_0+lf_{\mathrm{B}}$, but their amplitudes and phases follow shifted-harmonic and combination relations. At general non-linear order, coupling and modulation occupy the same frequency grid. In the current test samples, modulation-like classifications are obtained for 13/15 \textit{Kepler} and 9/10 K2 fundamental-mode RR Lyrae (RRab) stars, and for 4/6 TESS continuous-viewing-zone and 1/10 K2 first-overtone RR Lyrae (RRc) stars. Most inconclusive RRc cases, together with the two analysed high-amplitude $\delta$  Scuti stars, lack sufficient harmonic or common-side-peak support.}
   {The formalism provides a practical framework for interpreting Blazhko-like RRc stars, high-amplitude $\delta$  Scuti variables, and other pulsating stars in which modulation and coupling are difficult to separate observationally.  
   }
   \maketitle

\section{Introduction} 

Variations in the amplitudes and/or phases of light curves can be observed in several classes of pulsating stars. A prominent example is the Blazhko effect in RR Lyrae stars, where the pulsation amplitude and phase vary on a timescale that is much longer than the pulsation period. Despite this phenomenon having been discovered more than a century ago \citep{Blazhko_1907}, its physical origin remains unknown. For reviews of the Blazhko effect, see, for example, \citet{Smolec_Blazhko} and \citet{Kovacs_Blazhko2016}. 

Early ground-based observations suggested that some Blazhko stars show predominantly amplitude or predominantly phase variations \citep[e.g.][]{Kurtz2000ASPCS}. High-precision space photometry has shown, however, that amplitude and phase variations always occur simultaneously, although their relative strengths differ from star to star \citep{Szabo2014,Benko2014,Plachy2019,Molnar2022,Benko2023,Netzel2023}. 

From the observational point of view, the Blazhko effect manifests as a multiplet structure around the pulsation frequency and its harmonics in the Fourier spectrum. The most common interpretation of these multiplets is a single or multiple periodic modulation of a single oscillation. In this picture, the observed light curve can be described as the result of simultaneous amplitude modulation (AM) and frequency or phase modulation (FM/PM). A comprehensive mathematical formalism of this interpretation was developed by \citet{Benko2011} (hereafter B11), who showed how the observed multiplet structures naturally arise from periodic modulation and derived the expected relations between the amplitudes and phases of the Fourier components. \citet{Szeidl2012_modulation} pointed out that this simplified modulation formalism does not reproduce observed amplitudes and phases of real Blazhko stars. This apparent contradiction was later resolved by 
showing that the observed Blazhko variability need not be strictly periodic and is more generally represented by an almost-periodic formalism \citep{Benko2018}. 

An alternative interpretation of the observed amplitude and phase variation has been discussed in the literature. In the `beating' picture, the observed long-term variability is produced by the superposition of two or more oscillations with closely spaced frequencies. In its simplest form, the beating of two close independent frequencies can reproduce the apparent amplitude and phase variations. This has been demonstrated in detail for close-frequency pairs in $\delta$ Scuti stars \citep{BregerBischof2002,BregerPamyatnykh2006}, and a similar interpretation has been invoked for close secondary periodicities in RR Lyrae variables \citep{MoskalikPoretti2003}. 
More generally, non-linear interactions between the oscillations can generate additional combination frequencies and multiplet-like structures in the Fourier spectrum \citep{BregerKolenberg2006}.

It is very difficult to distinguish between modulation and beating based on observations: the two can produce similar amplitude and phase variations in the time domain and similar multiplet structures in the frequency domain. This issue has arisen in the case of several classes of pulsating stars. In first-overtone RR Lyrae stars (RRc variables), many Blazhko candidates exhibit highly asymmetric side frequencies, often consisting of side peaks that are detectable only on one side and thus form a doublet with the harmonics. Such doublet structures are commonly referred to as BL1, in contrast to BL2, triplet structures \citep{Alcock2003}. Such spectra can be interpreted either as signatures of strongly asymmetric modulation or as the consequence of an additional close frequency, whose beating with the dominant radial mode produces apparent long-term amplitude and phase variations \citep{Alcock2000,MoskalikPoretti2003,Netzel2018}.
Similar ambiguities occur in high-amplitude $\delta$ Scuti (HADS) stars, where long-term amplitude variations and multiplet structures have sometimes been discussed as possible Blazhko-like phenomena; the beating of close frequencies or non-linear mode interaction may provide alternative explanations \citep[e.g.][]{Breger2010vsgh.conf,Poretti2011,Bowman2018MNRAS,Yang_Esamdin2019}. 

The mathematical properties of modulation have been studied in considerable detail by B11.
Although \citet{Kurtz2015} presented a related unified Fourier description of non-sinusoidal stellar light curves in terms of a few base frequencies and their harmonics and combination frequencies, the explicit amplitude and phase relations for beating and non-linear coupling have not previously been developed at the
same level of generality. Consequently, it is not always clear which observed properties genuinely distinguish modulation from close-frequency beating or non-linear coupling, and
which are compatible with more than one interpretation. The aim of this study was to develop such a complementary formalism. Starting from a non-sinusoidal periodic signal and one or more nearby oscillations, we derived the Fourier spectra produced by linear superposition and by non-linear coupling. We compared the resulting frequency patterns, amplitudes, and phases with those expected from modulation, with the goal of identifying practical diagnostics that can be applied to observed stellar light curves.

\section{Beating of two close frequencies}\label{Sec:beating}

Throughout this paper we use the term `beating' in its strict sense, namely for the apparent amplitude and phase variations caused by the linear superposition of two or more close frequencies. We treated non-linear effects separately. When combination frequencies are generated from the interacting signals, we refer to this process as
`non-linear coupling' of close oscillations, but not as beating itself.

\subsection{Beating with a sinusoidal secondary component}

In the textbook sense, beating is the interference phenomenon produced by the linear superposition of two oscillations with nearby frequencies. Most textbooks discuss this using sinusoidal signals 
\citep[see e.g.][]{Feynman1963, French1971}.

Here, we chose as the primary signal the non-sinusoidal light curve of a variable star (e.g. RR Lyrae and HADS stars), which we expressed using its standard Fourier representation. Let us consider a periodic non-sinusoidal signal with fundamental frequency $f_0$, represented by a finite Fourier series
\begin{equation}
m_0(t)=A_0+\sum_{k=1}^{N}
A_k \sin \left( 2\pi k f_0 t+\varphi_k \right).
\label{eq:x0}
\end{equation}
\begin{figure}
    \centering
    \includegraphics[width=0.8\linewidth]{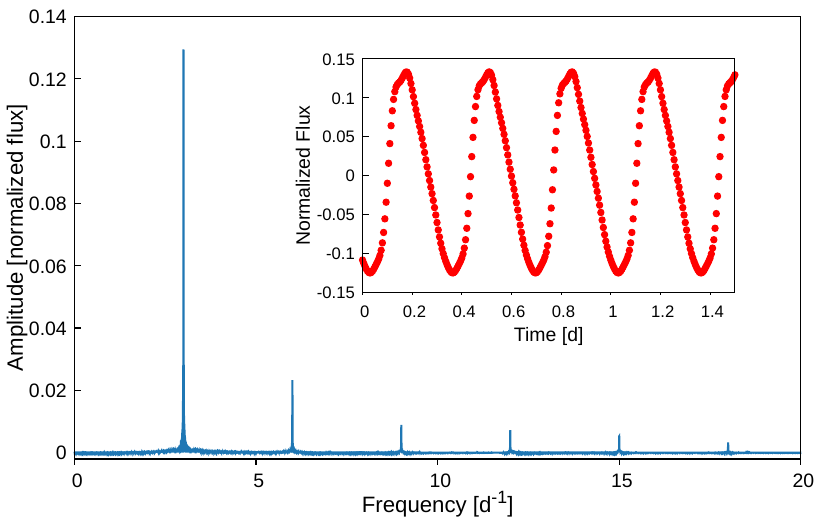}
    \caption{Fourier spectrum of the primary signal used in the generation of synthetic light curves, corresponding to Eq.~(\ref{eq:x0}) (\textit{main panel}), and a short segment of the light curve (\textit{inset}).
    }
    \label{fig:carrier}
\end{figure}
We generated the primary synthetic signal from the Fourier amplitudes and phases of the light curve of the RRc star NR Her obtained by the Transiting Exoplanet Survey Satellite (TESS), changing the main frequency to $f_0=3$~d$^{-1}$ for easier interpretation. The simulated time series span 200~d and have a sampling interval of 0.005~d. These settings produced high-quality synthetic light curves and Fourier spectra (see Fig.~\ref{fig:carrier}). 

In the beating framework, the observed variability is assumed to arise from the superposition of the $f_0$ signal and a second oscillation at a nearby higher frequency $f^{\prime}>f_0$. We defined $f_{\mathrm{B}}=f^{\prime}-f_0 \ll 1$~d$^{-1}$ as the beat frequency. Hereafter, we assumed $f^{\prime}>f_0$; the derivation for $f^{\prime}<f_0$ is analogous.

\begin{figure}
    \centering
    \includegraphics[width=0.8\linewidth]{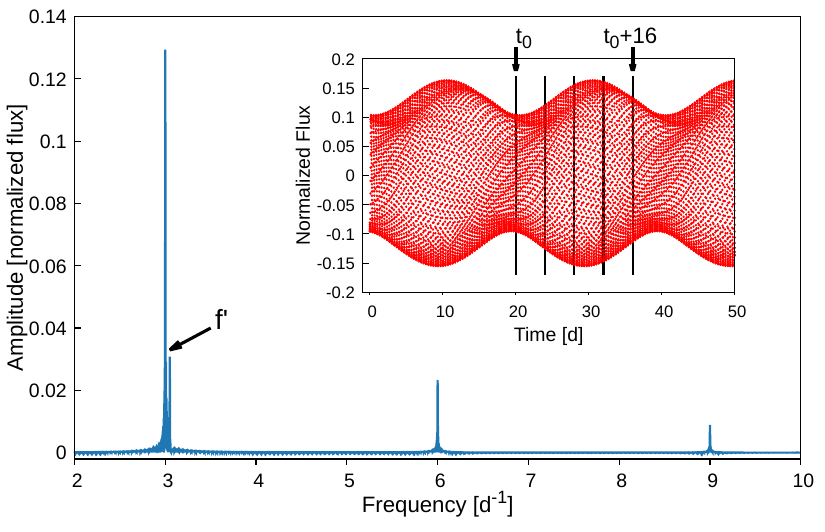}
    \includegraphics[width=0.8\linewidth]{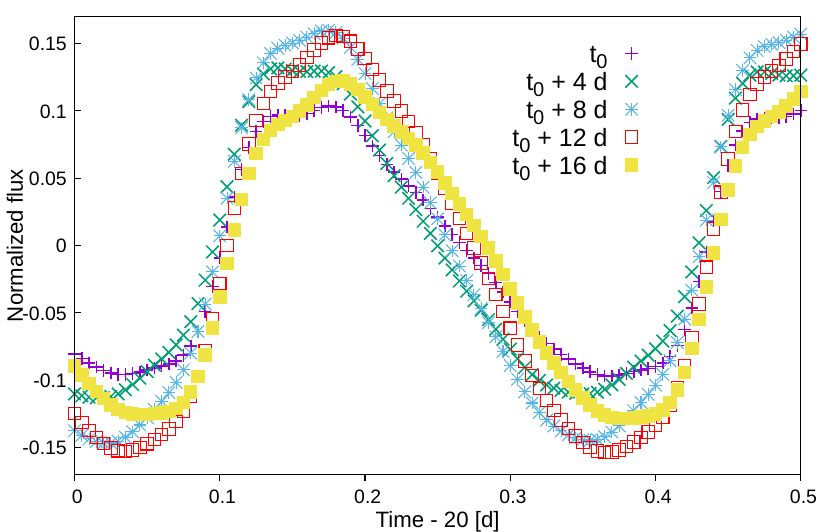}
    \caption{\textit{Top}: Fourier spectrum of an artificial light curve showing 20-day beating, with a sinusoidal secondary signal, according to Eq.~(\ref{eq:m0+m'}) (main panel), and a part of the light curve  (inset). \textit{Bottom}: 0.5-day segments extracted every four days and phase-folded using the same reference epoch, $t_0$. The vertical lines in the inset indicate the positions of the extracted light curve slices.
    }
    \label{fig:beat_sin}
\end{figure}

As the simplest case, let the second oscillation be purely sinusoidal,
\begin{equation}
m^{\prime}(t)=
B \sin \left(2\pi f^{\prime} t+\psi \right).
\label{eq:m'}
\end{equation}
The observed signal is
$ m(t)=m_0(t)+m^{\prime}(t)$ or
\begin{equation}
m(t)=A_0+B \sin[2\pi f^{\prime} t+\psi]+\sum_{k=1}^{N}
A_k \sin \left( 2\pi k f_0 t+\varphi_k \right).
\label{eq:m0+m'}
\end{equation}
Any constant term in the secondary signal can be absorbed into the zero point, $A_0$, of the observed light curve. We therefore took the secondary oscillation to have zero mean. Since Eq.~(\ref{eq:m0+m'}) corresponds to a Fourier representation, we can clearly see that, in addition to $f_0$ and its harmonics, only the frequency $f'$ appears. No further peaks are present (Fig.~\ref{fig:beat_sin}).

The first term of the series can be combined with Eq.~(\ref{eq:m'}) as
\begin{equation}
m(t)=A_0+C(t)\sin[2\pi f_0 t+\Phi(t)]+\sum_{k=2}^{N}
A_k \sin \left( 2\pi k f_0 t+\varphi_k \right),
\label{eq:beat_lin_t}
\end{equation}
where the expressions $C(t)$  and $\Phi(t)$ can be written as
\begin{equation}
C(t)= \sqrt{ A_1^2+B^2 +
2A_1B\cos(2\pi f_{\mathrm{B}} t+\psi-\varphi_1)},
\label{eq:C(t)}
\end{equation}
and
\begin{equation}
\tan\Phi(t) = \frac{ A_1\sin\varphi_1
+ B\sin(2\pi f_{\mathrm{B}} t+\psi)}
{A_1\cos\varphi_1 +
B\cos(2\pi f_{\mathrm{B}} t+\psi)}.
\label{eq:Phi(t)}
\end{equation}
As can be seen, only the main frequency participates directly in the beating. The harmonics of the primary oscillation remain unchanged in the Fourier spectrum. The synthetic light curve in Fig.~\ref{fig:beat_sin} was generated with $f'=3.05$~d$^{-1}$. The secondary signal produces a single peak near $f_0$, with no corresponding peaks near the higher harmonics
(main plot in the top panel). The shape of the non-sinusoidal light curve is distorted during beating (bottom panel).   

\begin{figure}
    \centering
    \includegraphics[width=0.49\linewidth, trim=0cm 2cm 0cm 0cm, clip]{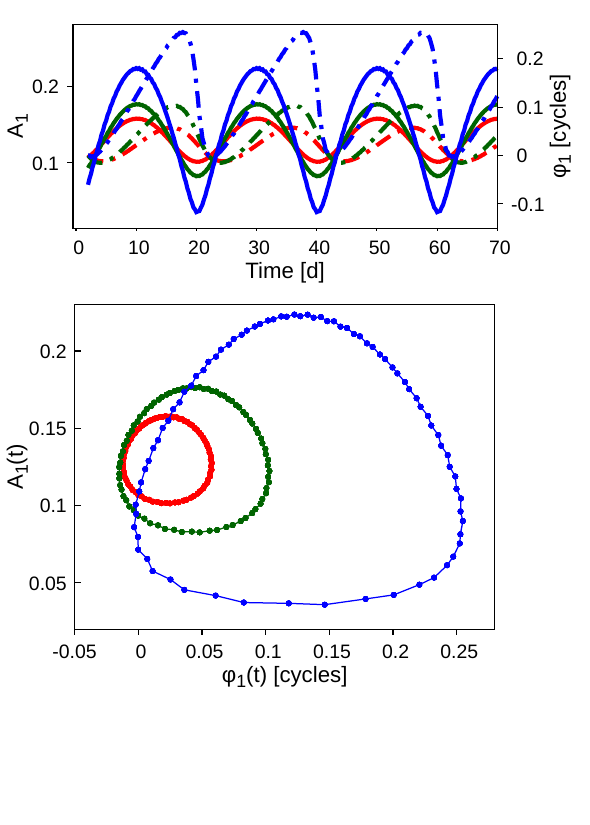}
    \includegraphics[width=0.49\linewidth, trim=0cm 2cm 0cm 0cm, clip]{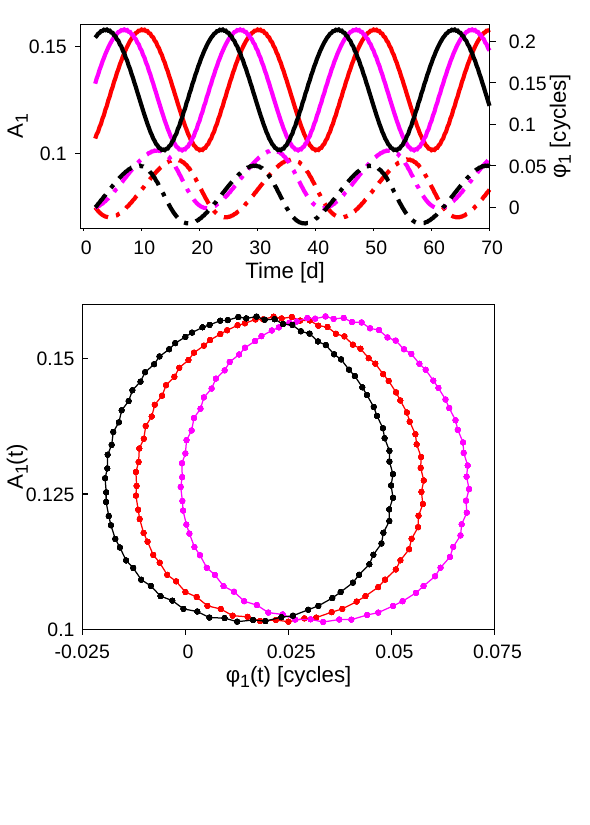}
    \caption{\textit{Top panels}: Variations in the amplitude (solid curves) and phase (dashed lines) of the main frequency of synthetic linear beating light curves. \textit{Bottom panels}: Their loop diagram. Here, we show a sinusoidal secondary signal. The secondary amplitudes are $B=0.03$ (red), $B=0.05$ (green), and $B=0.1$ (blue). The secondary phases are $\psi=1.2$ (red), $\psi=2.2$ (magenta), and $\psi=3.2$~rad (black).
    }
    \label{fig:beat_sin_loop}
\end{figure}

Figure~\ref{fig:beat_sin_loop} illustrates the amplitude and phase variations corresponding to Eqs.~(\ref{eq:C(t)}) and (\ref{eq:Phi(t)}) for several amplitude and phase values, as well as the loop diagrams often constructed from observational data. To simulate the analysis of observations, we derived the data in Fig.~\ref{fig:beat_sin_loop} from the artificial light curve using time-dependent Fourier analysis with a 4-day sliding window shifted in 0.25-day steps.
The observed $A_1(t)$ and $\varphi_1(t)$ functions correspond to the exact expressions of $C(t)$ and $\Phi(t)$. 

For small secondary amplitudes, $B \ll A_1$, Eqs.~(\ref{eq:C(t)}) and (\ref{eq:Phi(t)}) reduce to $C(t) \simeq A_1 + B\cos\theta$, and $\Phi(t) \simeq \varphi_1 + (B/A_1)\sin\theta$, where $\theta=2\pi f_{\mathrm{B}}t+\psi-\varphi_1$. In this limit, the amplitude and phase variations are shifted by approximately $\pi/2$ in beat phase, as shown by \citet{Bowman2016} and \citet{Bowman2018MNRAS} (see also Fig.~\ref{fig:beat_sin_loop}).
This is the quadrature behaviour often quoted as a signature of beating. The exact relation, however, is non-linear. For $B<A_1$ the extrema of $\Phi(t)$ occur at $\cos\theta=-B/A_1$, not exactly at $\theta=\pi/2$. As $B$ approaches $A_1$, the resultant amplitude minimum $C_{\min}=|A_1-B|\approx 0$ and the phase variation becomes poorly defined, producing the familiar apparent phase jump of approximately $\pi$ in a wrapped phase representation as previously shown by \citet{BregerPamyatnykh2006}. The transition between the two limiting cases is continuous.

\subsection{Beating with a non-sinusoidal secondary component}

\begin{figure}
    \centering
    \includegraphics[width=0.8\linewidth]{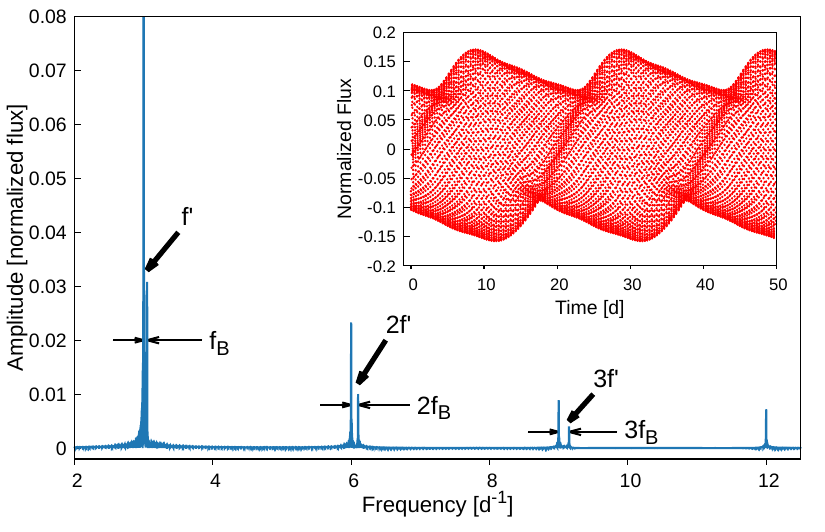}
    \caption{Fourier spectrum of a synthetic beating light curve with a non-sinusoidal secondary signal $f^{\prime}=3.05$~d$^{-1}$ containing three harmonics (\textit{main panel}), as well as a segment of the light curve (\textit{inset}). The side-peak separations from the harmonics are $lf_{\mathrm{B}}$ for $l=1, 2$, and 3.
    }
    \label{fig:beat_nonsin}
\end{figure}

If the secondary oscillation is also non-sinusoidal, it can be represented by a Fourier series,
\begin{equation}
m^{\prime}(t)= B_0 + \sum_{l=1}^{M} B_l \sin \left( 2\pi l f^{\prime} t+\psi_l \right).
\label{eq:nonlin}
\end{equation}
The observed signal is the sum of Eqs.~(\ref{eq:x0}) and (\ref{eq:nonlin}). Therefore,  the Fourier spectrum consists of two independent harmonic sequences $kf_0$ and $lf^{\prime}$ (see Fig.~\ref{fig:beat_nonsin}).
 
For the synthetic light curve, we included the first three harmonics of the secondary frequency $f^{\prime}$, corresponding to $l=1,2,3$. The inset in Fig.~\ref{fig:beat_nonsin} clearly shows that the envelope of the beating light curve is non-sinusoidal. 
The harmonics of the secondary oscillation appear at $ lf^{\prime} = lf_0+lf_{\mathrm{B}}$, including the first three harmonics $(l=1, 2, 3)$. Consequently, the separation from the corresponding primary harmonic, $lf_0$, increases linearly with harmonic order, $\Delta f_l=lf_{\mathrm{B}}$. This behaviour differs fundamentally from modulation, where the side-peak separation remains constant for all harmonic orders. In this example, no secondary peaks appear beyond $l=3$ (Fig.~\ref{fig:beat_nonsin}).
The amplitude and phase variations of the main frequency are the same as those of the sinusoidal case in Fig.~\ref{fig:beat_sin_loop}.

\begin{figure}
    \centering
    \includegraphics[width=0.49\linewidth, trim=0cm 2cm 0cm 0cm, clip]{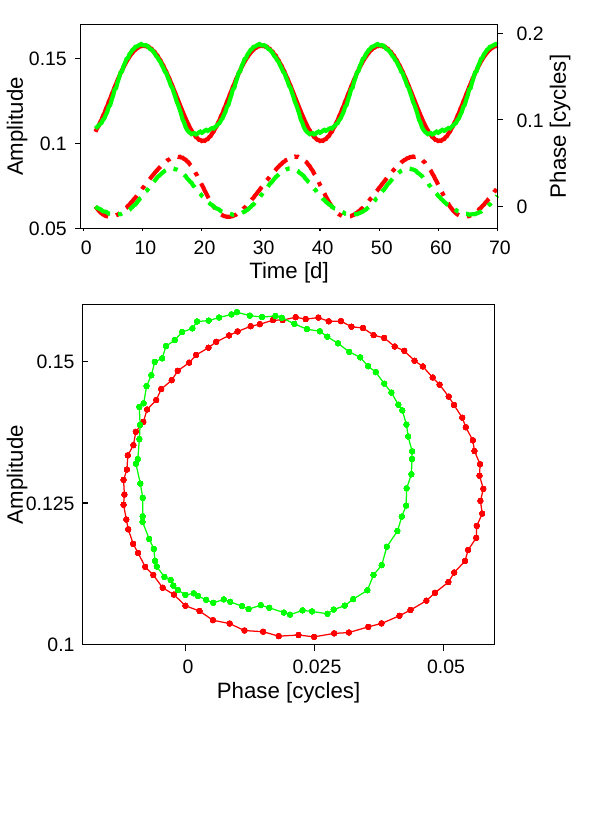}
    \includegraphics[width=0.49\linewidth, trim=0cm 2cm 0cm 0cm, clip]{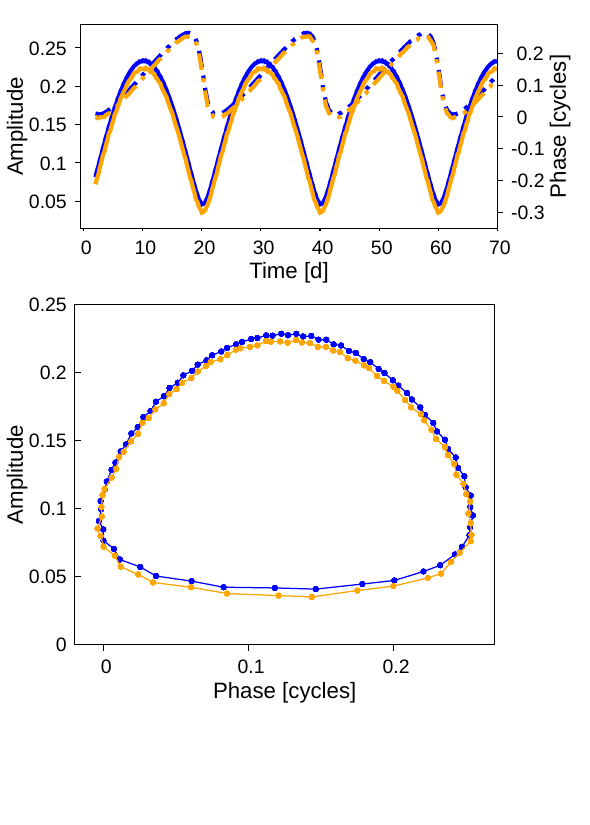}
    \caption{\textit{Top panels}: Variations over time in the main amplitudes $A_1(t)$ (solid~red and blue curves), half the total amplitudes (solid~green and orange curves), phases $\varphi_1(t)$ (dash-dotted~red and blue curves), and the total phases (dash-dotted~green and orange curves) of synthetic light curves where the secondary signal is non-sinusoidal. On the left, $B_1=0.03$; and on the right, $B_1=0.1$. \textit{Bottom panels}: Loop diagrams of the same light curves.
    (Orange curves have been shifted by +0.01 to make them easier to distinguish.)
    }
    \label{fig:beat_nonsin_loop}
\end{figure}

Of course, in the non-sinusoidal case, the amplitude and phase variations of the total light curve are not identical to the amplitude and phase variations of the main frequency, as they are in the sinusoidal case.
This is illustrated in Fig.~\ref{fig:beat_nonsin_loop}. The variations in the main frequency, $A_1(t)$ and $\varphi_1(t)$, are identical to those shown in Fig.~\ref{fig:beat_sin_loop} for the sinusoidal case. We determined the amplitude of the full light curve and the total phase using template fitting. 
In cases where $B_{\mathrm{tot}} \ll A_{\mathrm{tot}}$, both the amplitude and phase changes differ slightly from the sinusoidal case, but the loop diagram will be clearly different (left panels of Fig.~\ref{fig:beat_nonsin_loop}). At the same time, in cases where $B_{\mathrm{tot}} \approx A_{\mathrm{tot}}$, the variations are barely distinguishable (see the right panels of Fig.~\ref{fig:beat_nonsin_loop}).

\subsection{Multiple secondary signals}\label{Sec:multi_beating}

We considered the case in which more than one secondary oscillation is present. We write the observed light variation as 
\begin{equation} 
\begin{split}
m(t)=A_0+\sum_{k=1}^{N}
& A_k \sin \left( 2\pi k f_0 t  + \varphi_k \right) + \\
  & \sum_j \left[ B_{0,j}  + \sum_{l=1}^{M_j} B_{l,j} \sin \left( 2\pi l f_j^{\prime} t+\psi_{l,j} \right) \right],
\end{split}
\label{eq:linear_multiple} 
\end{equation}
where  $f^{\prime}_j=f_0+f_{\mathrm{B},j}$.  Equation~(\ref{eq:linear_multiple}) is a Fourier representation itself. Therefore, in the purely linear case the spectrum contains only the harmonics of the primary oscillation,  $k f_0$, and the harmonics of the secondary oscillations, $l f^{\prime}_j$. 
The latter can be written as $l f^{\prime}_j = l f_0+l f_{\mathrm{B},j}$. 
Thus, relative to the $l$th harmonic of the primary frequency, the $l$th harmonic of the $j$th secondary oscillation appears at a separation $ \Delta f_{l,j}=l f_{\mathrm{B},j}$. 

\begin{figure}
    \centering
    \includegraphics[width=0.8\linewidth]{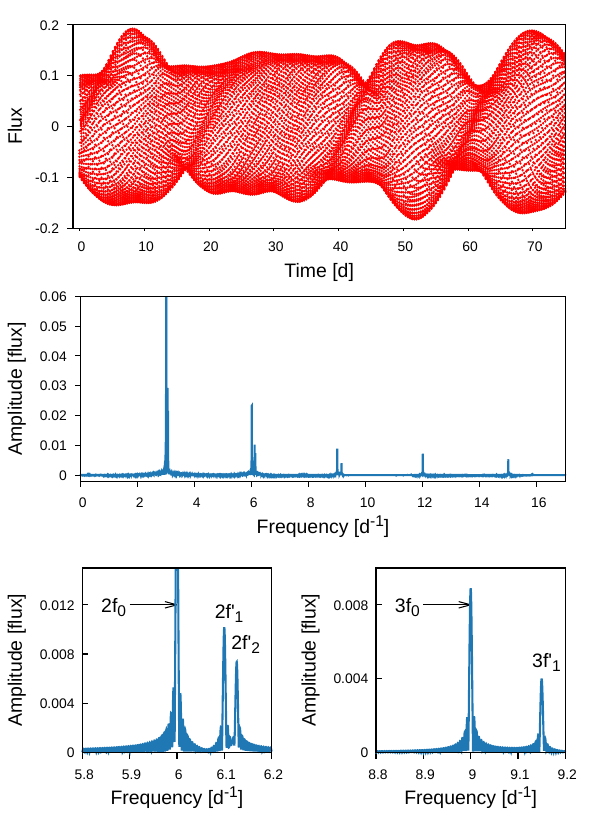}
    \caption{Segment of a synthetic beating light curve assuming non-sinusoidal secondary signals $f_1^{\prime}=3.05$~d$^{-1}$ and $f_2^{\prime}=3.063$~d$^{-1}$ with three and two harmonics, respectively (\textit{top}), the Fourier spectrum (\textit{middle}), and parts of it at around $2f_0$ and $3f_0$ (\textit{bottom}).
    }
    \label{fig:beat_multi}
\end{figure}

If several secondary frequencies are present, each of them generates its own harmonic sequence. Consequently, non-equidistant side-peak-like structures can already arise in the purely linear beating case (see also the bottom panels in Fig.~\ref{fig:beat_multi}). It is important to note that no difference frequencies such as $f_{\mathrm{B},j}=f^{\prime}_j-f_0$ are produced in this linear case. Similarly, there are no combination frequencies involving two secondary oscillations, such as $f^{\prime}_i\pm f^{\prime}_j$. 

\begin{figure}
    \centering
    \includegraphics[width=0.49\linewidth, trim=0cm 2cm 0cm 0cm, clip]{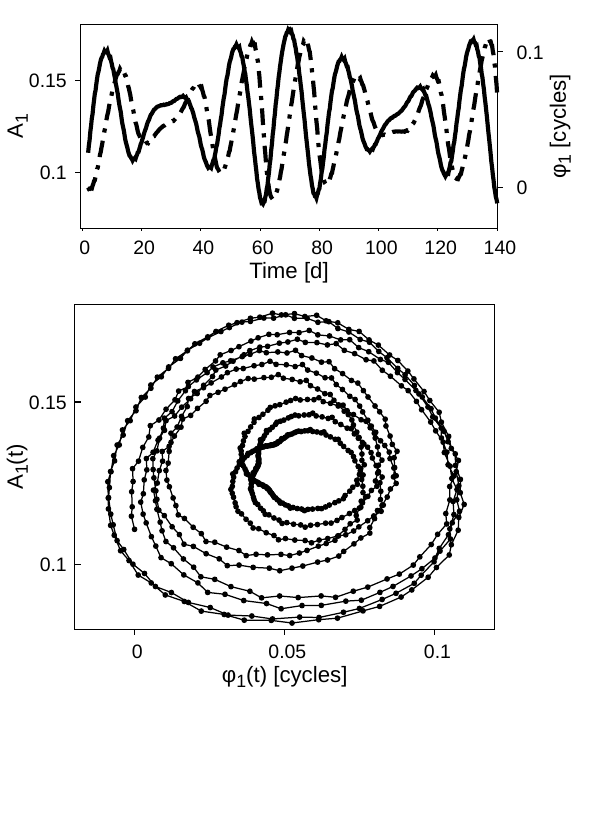}
    \includegraphics[width=0.49\linewidth, trim=0cm 2cm 0cm 0cm, clip]{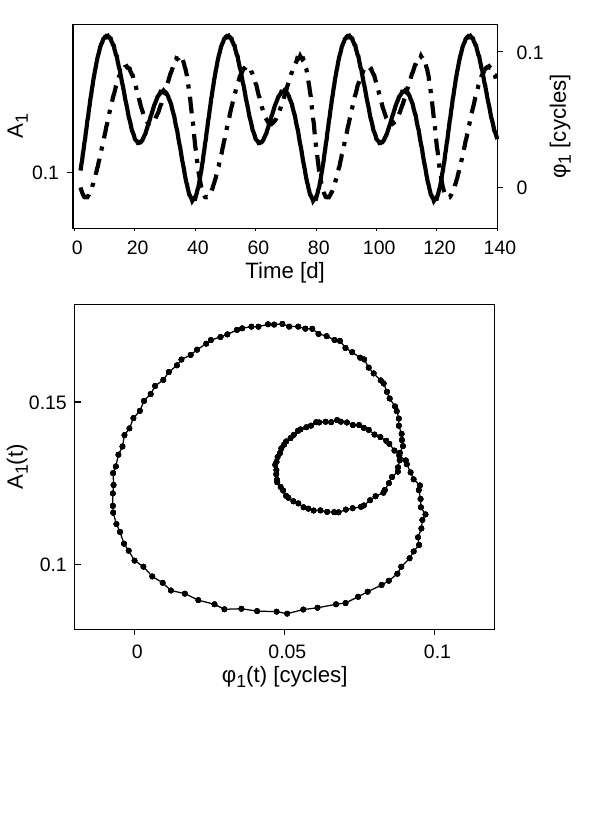}
    \caption{\textit{Top panels}: Variations in the amplitude (solid curves) and phase (dashed lines) of the main frequency of synthetic linear beating light curves. \textit{Bottom panels}: Their loop diagram. Here, we assumed two non-sinusoidal secondary signals. \textit{Left}:  $f_{\mathrm{B},2}=0.063$~d$^{-1}$. \textit{Right}: $f_{\mathrm{B},2}=0.025$~d$^{-1}$. In both cases, $f_{\mathrm{B},1}=0.05$~d$^{-1}$.
    }
    \label{fig:beat_multi_loop}
\end{figure}

If two components coincide exactly, for example if $l f^{\prime}_j = l' f^{\prime}_r $, for two different secondary oscillations, their complex Fourier amplitudes add vectorially (see Sect.~\ref{Sec:res} for the details). In practice, finite frequency resolution can also make two close but distinct components appear as a single broadened peak. Apart from such coincidences or resolution effects, however, the spectrum of Eq.~(\ref{eq:linear_multiple}) is simply the sum of the individual spectra.
The amplitude and phase variations of the main frequency are more complicated than in Fig.~\ref{fig:beat_sin_loop} or in Fig.~\ref{fig:beat_nonsin_loop}. In the general case, if the beat frequencies $f_{\mathrm{B},j}$ are not commensurable within the frequency resolution of the data, the resulting amplitude and phase variation functions are not periodic and the loop diagrams are open (Fig.~\ref{fig:beat_multi_loop}). 
For commensurate beat frequencies, such as $f_{\mathrm{B},1}=2f_{\mathrm{B},2}$, the amplitude and phase variations are periodic and the loop diagrams are closed (on the right side of Fig.~\ref{fig:beat_multi_loop}).

\section{Non-linear coupling}\label{sec:nonlin_coupl}

In the previous section, we discussed linear superposition. In real pulsating stars, however, the observed light variation is not expected to be a purely linear sum of independent sinusoidal oscillations. Stellar pulsation is intrinsically non-linear: the equations governing the pulsation contain non-linear terms
\citep[e.g.][]{Cox1980}, while the transformation from the underlying displacement, temperature, and radius variations to the observed flux variation is itself non-linear \citep[e.g.][]{Balona2012}. As a consequence, even if the star oscillates in two or more well-defined modes, the observed Fourier spectrum will contain not only the eigenfrequencies themselves, but also combination frequencies, unless they fall below the observational noise level.

Non-linear mode coupling has a long history in the theory of stellar pulsation. In the amplitude-equation formalism, non-linear terms describe the exchange of energy and phase information between pulsation modes, and they are essential for mode selection, amplitude limitation, resonant interactions, and frequency locking \citep[e.g.][]{Buchler_Goupil1984,Dziembowski1993}. In RR Lyrae stars, resonant interaction between the dominant radial mode and non-radial modes has been proposed as a possible route to Blazhko-type behaviour \citep{NowakowskiDziembowski2001, DziembowskiMizerski2004}. In rotating stars, pulsation and rotation can also interact: rotational splitting, near degeneracy, and spherical-harmonic mixing can modify the observed frequency pattern, especially when modes of different degree have close frequencies \citep[e.g.][]{Soufi1998}. Similar problems are also relevant for $\delta$ Scuti stars and HADS stars, where combination frequencies and non-linear mode interactions are commonly observed \citep[e.g.][and further references therein]{Breger2010vsgh.conf,Balona2016,Bowman2016}. 

In the present paper we use the term `non-linear coupling' in a phenomenological sense: the observed signal is assumed to be a non-linear function of a dominant non-sinusoidal radial pulsation and one or more additional close oscillations, which may be interpreted, for example, as non-radial modes.  
Thus, the formalism developed below should not be read as a specific physical theory of mode coupling, but as the mathematical consequence of allowing close oscillations to enter the observed light curve non-linearly.

\subsection{Quadratic coupling with a sinusoidal secondary signal}\label{Sec:quadratic_sin}

As the simplest non-linear case, let us consider quadratic coupling. Since the constant term of the primary signal does not generate a new combination frequency, it is convenient to introduce the mean-subtracted primary oscillation $\widetilde{m}_0(t)=m_0(t)-A_0$ and to write the model as
\begin{equation}
m(t)=m_0(t)+m^{\prime}(t)+\alpha \widetilde{m}_0(t)m^{\prime}(t).    
\label{eq:quadratic}
\end{equation}
This form is observationally equivalent, as far as the resulting frequency pattern is concerned, to using $m_0(t)$ in the product. The omitted term $\alpha A_0m^{\prime}(t)$ has exactly the same frequency content as the intrinsic secondary oscillation and can therefore be absorbed into its complex amplitude. 

It is often the case in observational data that the secondary oscillation has no harmonics, or that these cannot be detected, so first we discuss the sinusoidal secondary oscillation here.
Substituting Eqs.~(\ref{eq:x0}) and (\ref{eq:m'}) into Eq.~(\ref{eq:quadratic}), we obtain the non-linear term
\begin{equation}
\alpha \widetilde{m}_0(t)m^{\prime}(t) 
=\alpha\sum_{k=1}^{N}A_kB \sin(2\pi kf_0t+\varphi_k) \sin(2\pi f^{\prime}t+\psi).     
\end{equation}
Using the trigonometric formula for the product of sines,
\begin{equation}
\begin{split}
\alpha \widetilde{m}_0(t)m^{\prime}(t) = 
\frac{\alpha}{2} \sum_{k=1}^{N} A_kB \{\cos[ 2\pi ((k-1)f_0 - f_{\mathrm{B}})t+\varphi_k-\psi] - \\
\cos[ 2\pi ((k+1)f_0 + f_{\mathrm{B}})t+\varphi_k+\psi]\}.
\end{split}\label{eq:quad_lin_sin}
\end{equation}
Consequently, side peaks appear at $ nf_0\pm f_{\mathrm{B}}$ for $n=0,1,\dots, N+1$, subject to the index ranges specified below.

Equation~(\ref{eq:quad_lin_sin}) is written in terms of cosine functions, whereas the Fourier phases in Eqs.~(\ref{eq:x0})--(\ref{eq:m0+m'}) are defined using a sine convention. The corresponding $\pm\pi/2$ phase offsets must therefore be retained when the combination terms are expressed in the original sine convention. For a real coupling coefficient, all phases below are understood modulo $2\pi$, and $\arg\alpha$ is either 0 or $\pi$.

For the right-hand side peak at $nf_0+f_{\mathrm{B}}$,
\begin{equation}
A_n^{+}=\frac{|\alpha|}{2}A_{n-1}B, \quad
\phi_n^{+}=\varphi_{n-1}+\psi-\frac{\pi}{2}+\arg\alpha, \quad n=2,\ldots,N+1, 
\label{eq:A+}
\end{equation}
whereas for the left-hand side peak at $nf_0-f_{\mathrm{B}}$,
\begin{equation}
A_n^{-}=\frac{|\alpha|}{2}A_{n+1}B,\quad
\phi_n^{-}=\varphi_{n+1}-\psi+\frac{\pi}{2}+\arg\alpha, \quad n=1,\ldots,N-1.
\label{eq:A-}    
\end{equation}
Here we introduced $A_n^{\pm}= A(nf_0\pm f_{\mathrm{B}})$ and $\phi_n^{\pm}=\varphi(nf_0\pm f_{\mathrm{B}})$.
For $n=1$, the right-side peak is the secondary signal itself, at $f^{\prime}=f_0+f_{\mathrm{B}}$, so $A_1^+=B$ and $\phi_1^+=\psi$. The left-side peak is the combination peak given by Eq.~(\ref{eq:A-}).
The low-frequency difference term at $f_{\mathrm{B}}$, which originates from the $k=1$ term, should be treated separately because the formal expression $nf_0-f_{\mathrm{B}}$ has negative frequency for $n=0$. In the same sine convention,
\begin{equation}
A(f_{\mathrm{B}})=\frac{|\alpha|}{2}A_1B, \qquad
\phi(f_{\mathrm{B}})=\psi-\varphi_1+\frac{\pi}{2}+\arg\alpha.
\end{equation}

\begin{figure}
    \centering
    \includegraphics[width=0.8\linewidth]{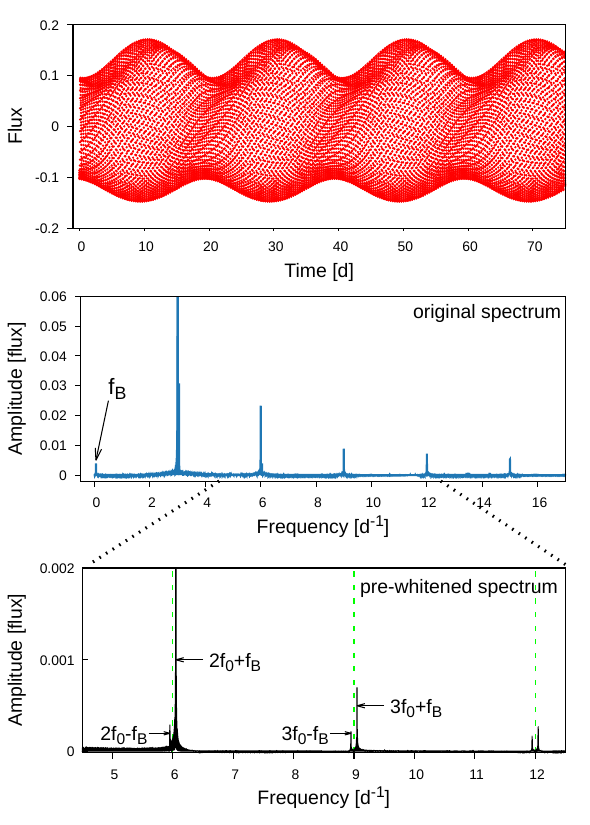}
    \caption{Segment of a synthetic light curve with a quadratically coupled sinusoidal secondary signal (\textit{top}), its Fourier spectrum (\textit{middle}), and a section of the residual spectrum after pre-whitening the main components (\textit{bottom}). The original and pre-whitened spectra are shown on different amplitude scales.
    }
    \label{fig:coupl_sin}
\end{figure}

Equation~(\ref{eq:quad_lin_sin}) and the corresponding Fourier spectrum were previously examined by \citet{Kolenberg_etal2006} in connection with the Blazhko effect of RR Lyrae itself.  This frequency structure 
 closely resembles a pure-AM spectrum. In the AM formalism, the modulation frequency $f_{\mathrm{m}}$ appears at low frequency if the mean level is also modulated. In the present quadratic-coupling model,
$f_{\mathrm{B}}$ arises naturally as a difference frequency (middle panel in Fig.~\ref{fig:coupl_sin}). 
The amplitudes, however, are completely different. First of all, the triplets produced by pure AM are always symmetric, meaning that the side frequencies belonging to the same order always have the same amplitude. Here, however, the triplets are generally asymmetric (see the bottom panel in Fig.~\ref{fig:coupl_sin}). The reason is that the side-frequency amplitudes are not proportional to the amplitude, $A_n$, of the corresponding primary harmonic but depend on neighbouring harmonics
(compare Eq.~(\ref{eq:quad_lin_sin}) with Eq.~(22) in B11).
In the case of AM, the side peaks appear at every harmonic, and only there. Here, the side peak on the right appears even one harmonic beyond the last one, while the side peak on the left no longer appears at the last harmonic (see the filled and empty green squares in Fig.~\ref{fig:ampl_lecseng}). 
These limits in harmonic order are falsifiable predictions of the quadratic coupling model, but the non-detection of a predicted component is informative only if that component should have been detectable at the local noise level.

\begin{figure}
    \centering
    \includegraphics[width=0.7\linewidth, trim=0cm 2cm 0cm 0cm, clip]{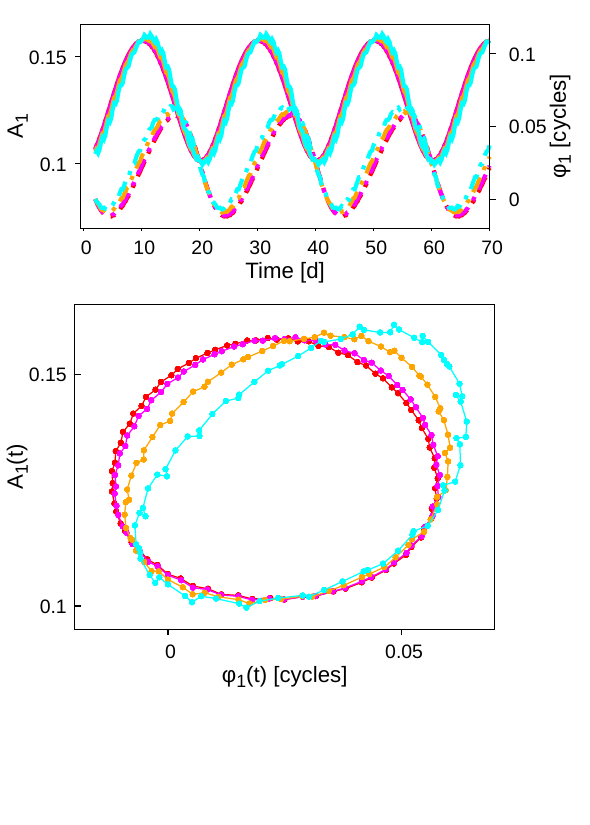}
    \caption{\textit{Top}: Variations in the amplitude (solid curves) and phase (dashed lines) of the main frequency of synthetic light curves showing quadratic coupling. \textit{Bottom}: Their loop diagram. The secondary signal is a simple sinusoidal function. The curves represent different coupling constants: red for $\alpha=0$ (the same as in Fig.~\ref{fig:beat_sin_loop}) and magenta, orange, and cyan for $\alpha=2, 10$, and 20, respectively.
    }
    \label{fig:loop_coupl_sin}
\end{figure}

If the additional combination peak at $f_0-f_{\mathrm{B}}$ has much lower amplitude than the intrinsic secondary component at $f^{\prime}=f_0+f_{\mathrm{B}}$, the time-dependent $A_1(t)$ and $\varphi_1(t)$ curves (Fig.~\ref{fig:loop_coupl_sin}) closely resemble the linear sinusoidal case shown in Fig.~\ref{fig:beat_sin_loop}. The coupling coefficient, $\alpha$, has no intrinsic upper limit in this phenomenological model; its numerical value depends on the normalisation of the light curve. However, for a weak quadratic coupling one should require $|\alpha|\,|\widetilde{m}_0(t)|\ll1$ and $|\alpha|\,|m'(t)|\ll1$ over the observed range. Larger values (e.g. $\alpha=20$ in Fig.~\ref{fig:loop_coupl_sin}) are mathematically allowed, but then the signal should be interpreted as an illustrative non-linear distortion rather than as a perturbative coupling term. Generally speaking, although the amplitude-phase loops of the coupled case never coincide with the diagram of the beating case, for physically relevant values of $\alpha\ne 0$, the amplitude, phase and loop diagrams are not robust discriminators between beating and quadratic coupling.

\subsection{Quadratic coupling with a non-sinusoidal secondary oscillation}\label{Sec:quadratic_nonsin}

From this point onwards, it is useful to adopt a complex representation. In place of Eqs.~(\ref{eq:x0}) and (\ref{eq:nonlin}), the primary and secondary signals are written as
\begin{equation}
x_0(t) = \sum_{k=-N}^{N} a_k \mathrm{e}^{\mathrm{i}k\omega_0 t},
\label{eq:x0_complex}
\end{equation}
and
\begin{equation}
x_1(t) = \sum_{l=-M}^{M} b_l \mathrm{e}^{\mathrm{i}l\omega_1 t},
\label{eq:x1_complex}
\end{equation}
where
$\omega_1=\omega_0+\Omega$, $\omega_0=2\pi f_0$, $\omega_1=2\pi f^{\prime}$ and $\Omega=2\pi f_{\mathrm{B}}$.
For real-valued signals, $a_{-k}=a_k^\ast$, $b_{-l}=b_l^\ast$. As usual, $\mathrm{i}$ denotes the imaginary unit, and the asterisks in the superscript denote complex conjugates.

Let us consider quadratic non-linear coupling,
\begin{equation}
X(t) = x_0(t) + x_1(t) + \alpha x_0(t)x_1(t).
\label{eq:quadratic_model}
\end{equation}
Substituting Eqs.~(\ref{eq:x0_complex}) and (\ref{eq:x1_complex}) into Eq.~(\ref{eq:quadratic_model}) yields
\begin{equation}
X(t) = \sum_k a_k \mathrm{e}^{\mathrm{i}k\omega_0 t} +
\sum_l b_l \mathrm{e}^{\mathrm{i}l\omega_1 t} +
\alpha \sum_{k,l} a_k b_l \mathrm{e}^{\mathrm{i}(k\omega_0+l\omega_1)t}.
\end{equation}
Using $ \omega_1=\omega_0+\Omega$, and introducing $n=k+l$,
the quadratic term becomes
\begin{equation}
\alpha \sum_{k,l}a_kb_l \mathrm{e}^{\mathrm{i}(k\omega_0+l\omega_1)t} =
\alpha \sum_{k,l}a_kb_l \mathrm{e}^{\mathrm{i}[n\omega_0+l\Omega]t}.
\end{equation}

The corresponding complex Fourier coefficient is therefore
\begin{equation}
C_{n,l} = a_n\delta_{l0} +
b_l\delta_{nl} +
\alpha a_{n-l}b_l ,
\label{eq:Cnm}
\end{equation}
where $a_j=0$ if $|j|>N$ and $b_j=0$ if $|j|>M$; $\delta_{nl}$ denotes the Kronecker delta. 
The three terms correspond respectively to
(i) the spectrum of the primary oscillation,
(ii) the spectrum of the secondary oscillation, and
(iii) quadratic combination frequencies.
The observable amplitude and phase are
\begin{equation}
A_{n,l}^{\mathrm{obs}} = 2|C_{n,l}|, \qquad \mathrm{and} \qquad \phi_{n,l}^{\mathrm{obs}} = \arg(C_{n,l}).
\label{eq:Aphiobs}
\end{equation}

\begin{figure}
    \centering
    \includegraphics[width=0.8\linewidth]{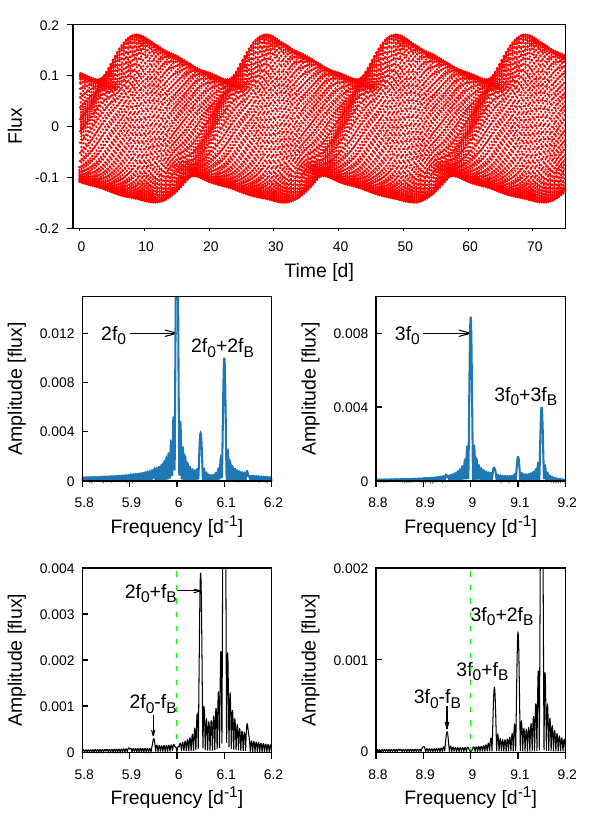}
    \caption{Segment of a synthetic light curve with a non-sinusoidal, quadratically coupled secondary signal at $f^{\prime}=3.05$~d$^{-1}$ containing three harmonics (\textit{top}), the Fourier spectrum at around $2f_0$ and $3f_0$ (\textit{middle}), and the same spectrum segments as in the middle but taken from the residual spectrum (\textit{bottom}). The vertical dashed lines indicate the positions of the pre-whitened harmonics.
    }
    \label{fig:coupl_nonsin}
\end{figure}

\begin{figure}
    \centering
    \includegraphics[width=0.8\linewidth]{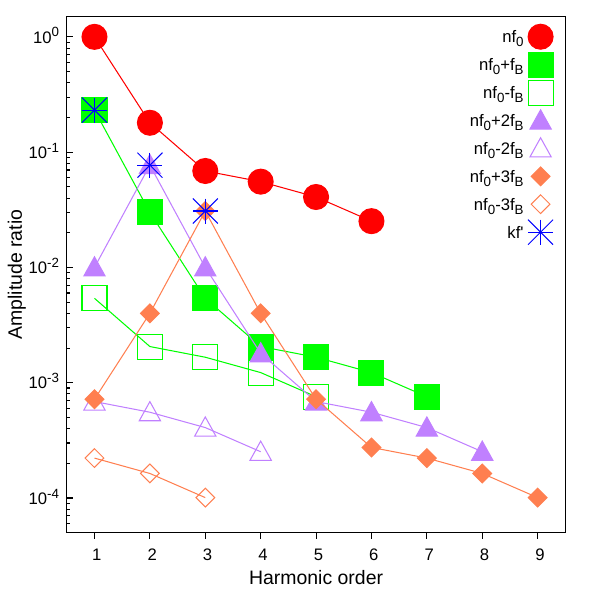}
    \caption{Amplitude ratios of a synthetic light curve by harmonic order, assuming a non-sinusoidal secondary signal. The vertical axis shows the amplitude ratios on a decimal logarithmic scale. The red circles show the primary harmonic amplitude ratios, $A(nf_0)/A(f_0)$, for $n=1,2,\ldots,6$. The green squares, purple triangles, and coral diamonds represent $A(nf_0\pm lf_{\mathrm{B}})/A(f_0)$, where $l=1,2,$ and 3, respectively. Filled and open symbols denote the upper and lower side peaks. The blue asterisks show the intrinsic harmonics of the secondary oscillation.
    }
    \label{fig:ampl_lecseng}
\end{figure}

For side peaks ($l\neq0$), Eq.~(\ref{eq:Cnm}) reduces to
\begin{equation}
C_{n,l} = b_l\delta_{nl} + \alpha a_{n-l}b_l .
\label{eq:side}
\end{equation}
If $n\neq l$ (i.e. we are not dealing with the position of the harmonic components of the secondary signal), only the combination term remains
\begin{equation}
C_{n,l} = \alpha a_{n-l}b_l .
\label{eq:purecomb}
\end{equation}
Consequently,
\begin{equation}
|C_{n,l}| = |\alpha|\,|a_{n-l}| \,|b_l|,
\label{Cnl}
\end{equation}
and
\begin{equation}
\arg(C_{n,l}) = \arg(\alpha) + \arg(a_{n-l}) + \arg(b_l).
\label{eq:phase_relation}
\end{equation}
Equation~(\ref{eq:purecomb}) clearly shows that the $l$th side frequency can appear in a multiplet only if the $l$th harmonic is also present in the secondary signal. For example, if the secondary signal contains only the first three harmonics ($l=1,2,3$), quadratic coupling cannot produce side peaks of order $|l|\geq4$ (see also the bottom panels in Fig.~\ref{fig:coupl_nonsin} and Fig.~\ref{fig:ampl_lecseng}). This is an important difference from modulation. If FM is also present, then, in principle, an infinite series of equidistant multiplets appears around every harmonic (see B11 for the details). The pattern of where side peaks appear around the harmonics is similar to what we saw in the sinusoidal case: the first-order right-side peak appears even at the $N+1$ harmonic, while the left-side peak appears only at $N-1$; for second-order peaks, they appear at $N+2$ and $N-2$, and so on (see Fig.~\ref{fig:ampl_lecseng}). 

Equation~(\ref{eq:phase_relation}) shows that the phases of the combination frequencies are not independent, but are constrained by the phases of the parent oscillations.
Using Eqs.~(\ref{Cnl}), (\ref{eq:phase_relation}) and (\ref{eq:Aphiobs}), the observable amplitude and complex-coefficient phase are
\begin{equation}
A_{n,l}^{\mathrm{obs}} = \frac{|\alpha|}{2}A_{n-l}B_l,
\qquad \mathrm{and} \qquad
\phi_{n,l}^{\mathrm{obs}} = \phi_{n-l}+\psi_l+\arg\alpha,
\label{eq:Aphiobs_2}
\end{equation}
which is a direct generalisation of Eqs.~(\ref{eq:A+}) and (\ref{eq:A-}).
Here, $\phi_{n-l}=\arg(a_{n-l})$ and $\psi_l=\arg(b_l)$ denote phases in the complex Fourier convention. Consequently, the fixed $\pm\pi/2$ offsets associated with converting between sine and complex-exponential representations are already contained in the definitions of $a_k$ and $b_l$. In the synthetic examples below we use $\alpha>0$, for which $\arg\alpha=0$.

The formula~(\ref{eq:Aphiobs_2}) provides a diagnostic relationship. Consider two side peaks with the same signed side order $l$:
\begin{equation}
    \phi_{n,l}^{\mathrm{obs}} -  \phi_{m,l}^{\mathrm{obs}} = \phi_{n-l} - \phi_{m-l}.
\end{equation}
In other words, the phase differences along a fixed signed-$l$ sequence are determined by the phases of the shifted primary harmonics. In fact, we can go even further. Let us define the combination phase parameter in the usual non-linear manner:
\begin{equation}
    \Phi_{n,l} =  \phi_{n,l}^{\mathrm{obs}} - \phi_{n-l} - \psi_l = {\mathrm{const}}.
\end{equation}
That is, if $\Phi_{n,l}$ is approximately constant over several independently measured components, the observed phases suggest a quadratic-coupling model. This is a very useful relationship, but it is not by itself a unique proof of quadratic coupling. The discriminatory power of the relation decreases if only a few components are available, or if the phase uncertainties are large. Coincident components of more general non-linear coupling (Sect.~\ref{Sec:general}) and sufficiently complex modulation models can also imitate a constant combination phase relation, although such a simple relation is generally not expected in these cases.

Figure~\ref{fig:ampl_lecseng} illustrates the progression of the harmonics and the side peaks according to their harmonic orders. For the synthetic example in Fig.~\ref{fig:ampl_lecseng}, we adopted $|\alpha|=2$ with $A_k$, $B_l$ < 1 for all $k$, and $l$. The side peaks are always smaller than the harmonics of the main oscillation. In contrast to modulation, however, it is striking that the amplitude profiles of the side peaks belonging to different orders are generally non-monotonic (see also Fig.~13 in B11). This is particularly noticeable in the higher side orders ($l>1$; see purple triangles and coral diamonds in Fig.~\ref{fig:ampl_lecseng}) and is a direct consequence of Eq.~(\ref{eq:Aphiobs_2}), which predicts that a side peak at harmonic order $n$ follows the amplitude progression of the shifted primary harmonics $A_{n-l}$. Therefore, the declining harmonic-amplitude progression of the primary light curve is mapped into the side-peak amplitudes. A pronounced non-monotonic progression of the higher-order side peaks (for low $n$) is consequently coupling-like, whereas a consistently declining side-order envelope is more naturally compatible with modulation. Modulation side peaks can also show non-monotonic behaviour but only at high $n$ (see B11).

For fundamental-mode RR Lyrae (RRab) stars, the harmonic amplitudes are known not to follow a simple monotonic exponential decrease; they show a reversal or a change in slope at higher harmonic orders \citep{Benko2016,Niu2026}. To our knowledge, an analogous systematic study has not yet been carried out for RRc stars. Nevertheless, the RRc light curve used for the present synthetic example (NR Her) also shows a break in its harmonic-amplitude progression. Although the amplitudes decrease monotonically, they are not well described by a single exponential slope: after the first few harmonics the decrease becomes noticeably shallower. This flattening of the high-order harmonic tail (red circles in Fig.~\ref{fig:ampl_lecseng}) is also reflected in the amplitudes of the quadratically coupled side peaks.

\subsection{Several independent secondary oscillations}\label{Sec:quad_multi_independent}

In the linear case discussed in Sect.~\ref{Sec:multi_beating}, several secondary oscillations already produce non-equidistant side-peak-like structures. However, the spectrum contains only the intrinsic harmonic sequences of the individual oscillations (see the bottom panels in Fig.~\ref{fig:beat_multi}).

We next examined how this picture changes if the secondary oscillations can be quadratically coupled to the primary signal; that is, we assumed that the primary oscillation interacts with several secondary oscillations whose frequencies are close to the main frequency,
\begin{equation}
x_j(t)= \sum_{l=-M_j}^{M_j} b_{j,l} \mathrm{e}^{\mathrm{i}l\omega_j t},
\qquad
\omega_j=\omega_0+\Omega_j,
\qquad j=1,\ldots,J .
\end{equation}
The total secondary signal is
\begin{equation}
x_{\mathrm{s}}(t)=\sum_{j=1}^{J}x_j(t).
\label{eq:multi_sec}
\end{equation}

As a first step, we assumed that the secondary oscillations do not interact with one another and that only the coupling between the primary oscillation and each secondary oscillation is present,
\begin{equation}
X(t)=x_0(t) + x_{\mathrm{s}}(t) + \alpha x_0(t)x_{\mathrm{s}}(t).
\label{eq:independent}
\end{equation}
Substituting Eq.~(\ref{eq:multi_sec}) into Eq.~(\ref{eq:independent}) yields
\begin{equation}
X(t)=x_0(t) + \sum_j x_j(t) + \alpha \sum_j x_0(t)x_j(t).
\end{equation}
Each term $x_0x_j$ produces frequencies $(k+l)\omega_0+l\Omega_j$.
Introducing $n=k+l$, the observable frequencies become
$n\omega_0+l\Omega_j$, or, equivalently,
\begin{equation}
f_{j,nl}=nf_0+l f_{\mathrm{B},j},
\qquad{\mathrm{where}}\qquad
f_{\mathrm{B},j}=f^{\prime}_j-f_0 .
\label{eq1}
\end{equation}
Thus, each secondary oscillation generates its own multiplet structure around every harmonic of the primary oscillation, $nf_0$, $nf_0\pm f_{\mathrm{B},j}$, $nf_0\pm2f_{\mathrm{B},j},\ldots$. Equation~(\ref{eq1}) shows that the superposition of several close frequencies naturally produces distinct multiplet systems. If the beat frequencies are not equal, the resulting side peaks need not be equidistant around the harmonics of the primary oscillation.

The complex Fourier coefficient is
\begin{equation}
C_{n,l}^{(j)}=b_{j,l}\delta_{nl} + \alpha a_{n-l}b_{j,l}.
\end{equation}

\noindent Away from the harmonics of the secondary oscillation ($n\neq l$),
\begin{equation}
C_{n,l}^{(j)}=\alpha a_{n-l}b_{j,l}.
\end{equation}
Consequently,
\begin{equation}
A_{n,l}^{{\mathrm{obs}} (j)}=\frac{|\alpha|}{2} A_{n-l}B_{j,l}, \quad \mathrm{and}\quad \phi_{n,l}^{{\mathrm{obs}} (j)}=\phi_{n-l} + \psi_{j,l}+\arg\alpha,
\end{equation}
assuming a real coupling coefficient.

\begin{figure*}
    \centering
    \includegraphics[width=0.4\linewidth]{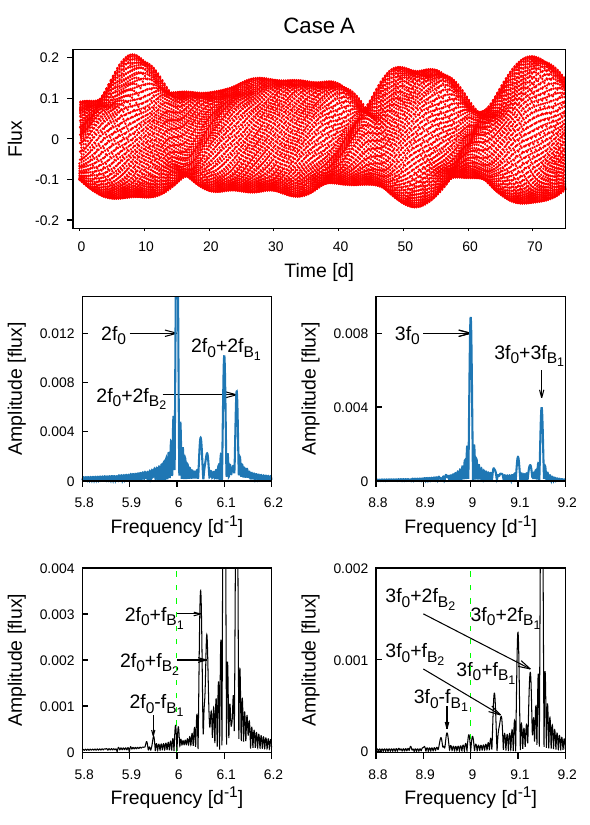}
    \hspace{0.05\columnwidth}
    \includegraphics[width=0.4\linewidth]{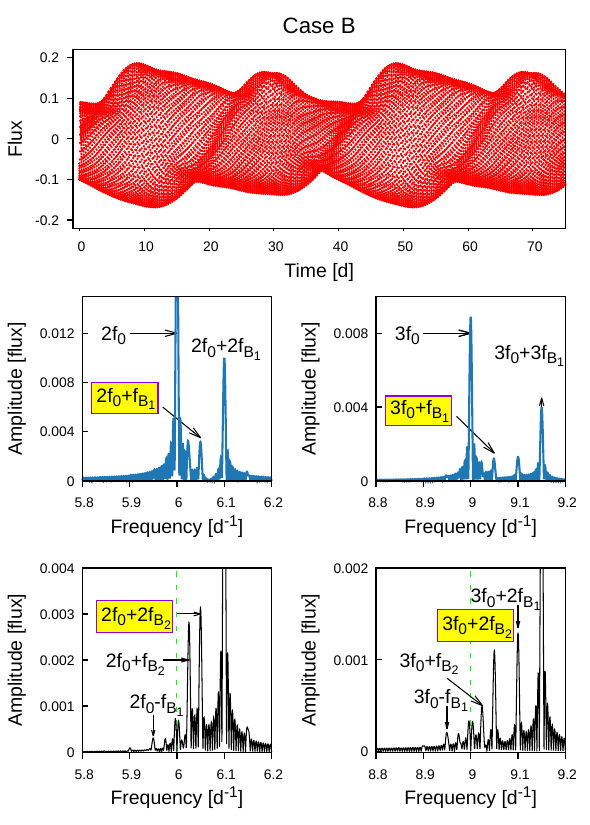}
    \caption{\textit{Top panels}: Synthetic light-curve segments containing two non-sinusoidal secondary signals, $f^{\prime}_1$ and $f^{\prime}_2$, each quadratically and independently coupled to the primary oscillation. The secondary signals contain three and two harmonics, respectively. \textit{Middle panels}: Fourier spectra around $2f_0$ and $3f_0$. \textit{Bottom panels}: Same frequency ranges after pre-whitening the primary harmonics. The dashed vertical lines mark the positions of the pre-whitened harmonics. In Case A, $f^{\prime}_1=3.05~\mathrm{d}^{-1}$ and $f^{\prime}_2=3.063~\mathrm{d}^{-1}$; the corresponding beat-frequency offsets have no low-order commensurability within the frequency resolution. In Case B, $f^{\prime}_1=3.05~\mathrm{d}^{-1}$ and $f^{\prime}_2=3.025~\mathrm{d}^{-1}$, so $f_{\mathrm{B},1}=2f_{\mathrm{B},2}$. Frequencies at which different contributions coincide in Case B are highlighted by yellow boxes.
    }
    \label{fig:coupl_multi}
\end{figure*}

The light curve itself is phenomenologically very similar to that of the linear case (cf. Fig.~\ref{fig:beat_multi} and Case A in Fig.~\ref{fig:coupl_multi}). However, the fine structure of the spectra reveals the presence of non-linear coupling (see the middle and bottom panels of Figs.~\ref{fig:beat_multi} and \ref{fig:coupl_multi}).
As in the single-secondary case discussed in Sect.~\ref{Sec:quadratic_sin}, the amplitude and phase variations and the loop diagrams closely resemble those of the corresponding linear beating model (as in Fig.~\ref{fig:beat_multi_loop}).

\subsubsection{The special case, $n=l$}\label{Sec:res}

The previous expressions were derived for frequencies that do not coincide with the harmonics of the secondary oscillations ($n\neq l$). The case $n=l$ requires special attention because the intrinsic harmonics of the secondary oscillations occur at the same frequencies.

For a single secondary oscillation, Eq.~(\ref{eq:Cnm}) gives
\begin{equation}
C_{l,l} = b_l + \alpha a_0 b_l .
\end{equation}
If we subtract the mean from the primary oscillation ($a_0=0$), this reduces to
\begin{equation}
C_{l,l}=b_l .
\end{equation}
Thus, in the single-secondary-frequency case, the harmonics of the secondary oscillation remain unaffected by the quadratic coupling, $A_{l,l}^{\mathrm{obs}}=B_l$ and $\phi_{l,l}^{\mathrm{obs}}=\psi_l$.

The situation becomes more complicated when more than one independent secondary oscillation is present. The frequency associated with the $l$th harmonic of the $j$th oscillation is
\begin{equation}
l f^{\prime}_j = l f_0+l f_{\mathrm{B},j}.
\end{equation}
A combination frequency generated by another oscillation $r$ can
coincide with this frequency if
\begin{equation}
l' f_{\mathrm{B},r} = l f_{\mathrm{B},j}.
\label{eq:coincidence_condition}
\end{equation}
In this case several contributions appear at the same Fourier frequency and their complex amplitudes must be added vectorially. From an observational standpoint, a `near coincidence' is also considered a coincidence when $\vert l' f_{\mathrm{B},r} - l f_{\mathrm{B},j}\vert \lesssim 1/T$, where $T$ is the time span of the time series.

Defining $\Lambda=l\Omega_j$, the total complex Fourier coefficient at the frequency
$\omega=n\omega_0+\Lambda$ is
\begin{equation}
C(n,\Lambda) = \sum_{\substack{j,l\\ l\Omega_j=\Lambda}}
\left[b_{j,l}\delta_{nl} + \alpha a_{n-l}b_{j,l} \right].
\label{eq:multi_C}
\end{equation}
The observable amplitude and phase are therefore
\begin{equation}
A^{\mathrm{obs}}(n,\Lambda) = 2|C(n,\Lambda)|, \quad \mathrm{and} \quad 
\phi^{\mathrm{obs}}(n,\Lambda) = \arg\!\left[C(n,\Lambda)\right] .
\end{equation}

If no coincidence satisfying Eq.~(\ref{eq:coincidence_condition})
exists, Eq.~(\ref{eq:multi_C}) reduces to
\begin{equation}
C(l,l\Omega_j) = b_{j,l} + \alpha a_0 b_{j,l}.
\end{equation}
For $a_0=0$, $C(l,l\Omega_j)=b_{j,l}$, which is identical to the single-secondary-frequency case.

However, if Eq.~(\ref{eq:coincidence_condition}) is satisfied,
the observed coefficient becomes
\begin{equation}
C(l,l\Omega_j) = b_{j,l} +
\sum_{\substack{(r,l')\neq(j,l)\\
l'\Omega_r=l\Omega_j}}
\alpha a_{l-l'} b_{r,l'} ,
\label{eq:res}
\end{equation}
showing that the intrinsic harmonic of one secondary oscillation can be modified by combination terms originating from another secondary oscillation. Consequently, in the presence of multiple secondary oscillations, the harmonics of a given oscillation are no longer guaranteed to appear as isolated Fourier peaks. Their amplitudes and phases can be altered by constructive or destructive interference with combination frequencies generated by the other oscillations. 

This behaviour is illustrated by the two spectra in Fig.~\ref{fig:coupl_multi}. In Case A, the two beat-frequency offsets are not commensurate at low integer order within the frequency resolution of the data. The combined amplitude and phase variations therefore do not repeat on a single beat period, and the light-curve envelope appears irregular over the plotted interval. In Case B, $f_{\mathrm{B},1}=2f_{\mathrm{B},2}$, so the variability has the common period $1/f_{\mathrm{B},2}$. In addition, the relation $f_{\mathrm{B},1}=2f_{\mathrm{B},2}$ causes several otherwise distinct non-linear contributions to fall at the same Fourier frequencies. Their complex amplitudes must then be added vectorially, producing constructive interference at some frequencies and destructive interference at others. Comparing the middle and bottom panels of Fig.~\ref{fig:coupl_multi}, for example, the amplitude of the coincident frequency $2f_0+f_{\mathrm{B},1}=2f_0+2f_{\mathrm{B},2}$ is smaller than the amplitudes of the non-coinciding frequencies taken separately from Case A. In contrast, the amplitude of the coincident frequency of $3f_0+f_{\mathrm{B},1}=3f_0+2f_{\mathrm{B},2}$ is higher than the amplitudes of the frequencies measured separately in Case A. These examples clearly illustrate the implication of Eq.~(\ref{eq:res}).

The coincidence condition in Eq.~(\ref{eq:coincidence_condition}) is the Fourier analogue of a resonance condition. In the present formalism, resonance means that two otherwise different frequency combinations fall at the same Fourier frequency, or at frequencies that cannot be resolved within the observational time base. This distinction is important because the term `resonance' is used in several, partly different, senses in the literature. 

In a physical mode-coupling sense, such a frequency relation can indicate an actual exchange of energy and phase information between modes. This is a stronger statement than the mere presence of a combination frequency. In $\delta$ Scuti stars, both interpretations have been discussed extensively. Combination frequencies can arise from non-linear light-curve distortion or from the non-linear transformation between the stellar displacement and the observed flux, but near-resonant mode coupling can also amplify otherwise weak components and impose characteristic amplitude and phase relations \citep[e.g.][]{Breger_Montgomery2014, BarceloForteza2015, Balona2016,MourabitWeinberg2023}. 

The formalism developed here does not distinguish by itself between these physical possibilities. It only specifies the frequency conditions under which different contributions coincide and how their complex amplitudes must be combined. Observationally, the diagnostic information is therefore not only the frequency coincidence itself, but also whether the amplitudes and phases of the involved peaks satisfy the expected combination or mode-coupling relations.

\subsection{General quadratic coupling}\label{Sec:quadratic_multi_general}

The previous model in Sect.~\ref{Sec:quad_multi_independent} assumes that the secondary oscillations are
mutually independent. In a more general quadratic description, all oscillations are allowed to interact with one another.

Introducing $x_0(t),x_1(t),x_2(t),\ldots,x_J(t)$, the most general quadratic model can be written as
\begin{equation}
X(t)=\sum_{p=0}^{J}x_p(t) 
+ \sum_{0\le p\le q\le J} \alpha_{pq}x_p(t)x_q(t) .
\label{eq:general_quadratic}
\end{equation}
The terms $x_0x_j$ reproduce all the frequency structures discussed above. However, the cross-coupling terms
$x_i(t)x_j(t)$, $(i\neq j)$, generate additional frequencies.

For two secondary oscillations with fundamental angular frequencies $\omega_i=\omega_0+\Omega_i$,
$\omega_j=\omega_0+\Omega_j,$
the product of their fundamental components generates a sum frequency and a difference frequency,
$\omega_i+\omega_j=2\omega_0+\Omega_i+\Omega_j$, and $|\omega_i-\omega_j|=|\Omega_i-\Omega_j|$.
In cyclic-frequency units, these components occur at
\begin{equation}
2f_0+f_{\mathrm{B},i}+f_{\mathrm{B},j}, \qquad \mathrm{and} \qquad |f_{\mathrm{B},i}-f_{\mathrm{B},j}|.
\end{equation}
Thus, the sum of the two beat-frequency offsets does not appear as an isolated low-frequency component: it remains attached to the $2f_0$ term. The genuinely low-frequency quadratic contribution is the difference $|f_{\mathrm{B},i}-f_{\mathrm{B},j}|$.

For non-sinusoidal secondary oscillations, the more general cross terms have frequencies
\begin{equation}
l f'_i \pm m f'_j
= (l\pm m)f_0+l f_{\mathrm{B},i}\pm m f_{\mathrm{B},j}.
\end{equation}
A low-frequency term is obtained only when the contributions proportional to $f_0$ cancel. For the difference of equal harmonic orders ($l=m$), this gives $l|f_{\mathrm{B},i}-f_{\mathrm{B},j}|$. General secondary--secondary quadratic coupling therefore introduces two observationally distinct families: low-frequency differences between the secondary offsets and additional sum or difference combinations around non-zero multiples of $f_0$.

\begin{figure}
    \centering
    \includegraphics[width=0.8\linewidth]{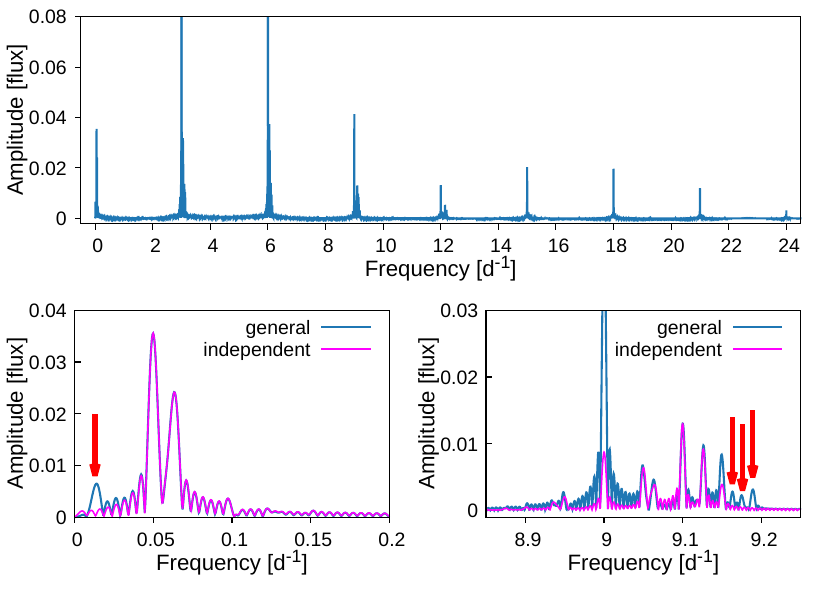}
    \caption{\textit{Top}: Fourier spectrum of a synthetic light curve containing two quadratically coupled non-sinusoidal secondary signals $f_1^{\prime}=3.05$~d$^{-1}$ with three harmonics and  $f_2^{\prime}=3.063$~d$^{-1}$ with two harmonics. \textit{Bottom}: Comparison of two Fourier spectra in the vicinity of $0$ and $3f_0$. The blue spectrum illustrates the general quadratic coupling (Sect.~\ref{Sec:quadratic_multi_general}), and the magenta spectrum shows the independent primary--secondary coupling case of Sect.~\ref{Sec:quad_multi_independent}, in which the secondary oscillations do not couple to one another. The arrows indicate the additional frequencies that appear only in the general-coupling case.
    }
    \label{fig:coupl_gen}
\end{figure}

The simultaneous detection of a low-frequency difference term and the corresponding cross-combination peaks near the pulsation harmonics would support direct interaction between the secondary oscillations. 
This is clearly illustrated in Fig.~\ref{fig:coupl_gen}, where we compare the spectra of the two quadratically coupled synthetic light curves. The input parameters of these light curves were identical; only the nature of the coupling differs between the two spectra. 
For example, the general quadratic model produces a difference-frequency component at $|f_{\mathrm{B},2}-f_{\mathrm{B},1}|$ that is absent from the independent primary--secondary model (bottom-left panel of Fig.~\ref{fig:coupl_gen}).
In the general quadratic case the self-coupling term \(x_0^2\) also contributes to the central harmonic frequencies \(n f_0\). Therefore, the amplitudes of the main harmonics are no longer
identical to the input amplitudes of the uncoupled primary signal (especially for the deliberately large coupling coefficient, $\alpha=20$, used in Fig.~\ref{fig:coupl_gen}). 
This is clearly seen when comparing the blue and magenta $3f_0$ peaks in the bottom-right panel of Fig.~\ref{fig:coupl_gen}. 
In this general quadratic-coupling example, the harmonic-amplitude progression shows a reversal similar to that observed in RRab stars (top panel in Fig.~\ref{fig:coupl_gen}).

\subsubsection{Relation to three-mode resonances} 

In the stellar-pulsation literature, three-mode sum resonances are commonly written in the form
\citep[e.g.][]{Dziembowski1982, VanBeeck2024}
\begin{equation}
   \omega_q \simeq \omega_r+\omega_s,
\end{equation}
or, equivalently, \(f_q \simeq f_r+f_s\), with a small de-tuning. In the present formalism the product \(x_r(t)x_s(t)\) generates Fourier components at the sum and difference frequencies \(\omega_r+\omega_s\) and \(|\omega_r-\omega_s|\). If an intrinsic oscillation is also present at a frequency \(\omega_q\) close to one of these combinations, then the intrinsic component and the non-linear combination term can contribute to the same observed Fourier frequency. For example, if 
$\Delta_{qrs}=\omega_q - \omega_r - \omega_s$
satisfies \(|\Delta_{qrs}|\lesssim 2\pi/T\), where \(T\) is the time span of the data, the two contributions are not resolved observationally. In that case the observed complex amplitude at \(\omega_q\) is a vector sum of the intrinsic mode amplitude and the quadratic combination contribution, 
\begin{equation}
   C^{\mathrm{obs}}(\omega_q) \simeq a_q+\alpha_{rs}a_ra_s , 
\end{equation} 
apart from convention-dependent normalisation and phase offsets. If the de-tuning is larger than the frequency resolution, the intrinsic mode and the combination term appear as a close frequency pair, producing an additional beating pattern on the de-tuning timescale. This frequency coincidence should not by itself be identified with a dynamical resonance. The present formalism specifies the Fourier frequencies, amplitudes, and phases expected from a non-linear transformation of the oscillations into the observed light curve. 

\section{General non-linear coupling}\label{Sec:general}

\subsection{One secondary signal}

Let the observed signal be an arbitrary analytic function of the two oscillations, 
\begin{equation} X(t)=F[x_0(t),x_1(t)] . 
\end{equation} 
Expanding $F$ into a power series gives 
\begin{equation} F(x_0,x_1) = \sum_{p,q\geq 0} c_{pq} x_0^p x_1^q . 
\end{equation} 
Using the complex Fourier representations of $x_0(t)$ and $x_1(t)$ given in Eqs.~(\ref{eq:x0_complex}) and (\ref{eq:x1_complex}), each non-linear term generates frequencies of the form $ K\omega_0+L\omega_1$. These frequencies can be rewritten as 
\begin{equation} K\omega_0+L\omega_1 = (K+L)\omega_0+L\Omega . 
\end{equation} 
Introducing  $n=K+L$, the peaks therefore appear at 
$\omega_{n,L}=n\omega_0+L\Omega$ , $n=0,\pm1,\pm2,\ldots$, $L=0,\pm1,\pm2,\ldots$. 
In frequency units this becomes 
\begin{equation} f_{n,L}=nf_0+Lf_{\mathrm{B}} . 
\label{eq:general_frequency_grid} 
\end{equation} 
Thus, an arbitrary non-linear coupling between two close oscillations naturally produces a frequency grid of the same formal type as the multiplet structure usually associated with modulation. In particular, around the $n$th harmonic of the primary frequency, peaks can appear at  $nf_0$, $nf_0\pm f_{\mathrm{B}}$, $nf_0\pm2f_{\mathrm{B}}$, $nf_0\pm3f_{\mathrm{B}},\ldots$. 

This grid represents only the set of allowed frequencies. The actual presence of a given multiplet component depends on both the harmonic content of the secondary oscillation and the order of the non-linear coupling. Consequently, the finite side-order restriction derived for quadratic coupling no longer applies: 
side peaks of order higher than the detectable harmonics of the secondary signal can also occur.

The complex Fourier coefficient at the frequency $n\omega_0+L\Omega$ is obtained by summing all non-linear paths that lead to the same frequency. For the term $c_{pq}x_0^p x_1^q$ this gives 
\begin{equation} C_{n,L}^{(p,q)} = c_{pq} \sum_{\substack{ k_1+\cdots+k_p+l_1+\cdots+l_q=n\\ l_1+\cdots+l_q=L }} a_{k_1}\cdots a_{k_p} b_{l_1}\cdots b_{l_q}. 
\end{equation} 
The total complex coefficient is therefore 
\begin{equation} C_{n,L} = \sum_{p,q\geq0} C_{n,L}^{(p,q)} , 
\label{eq:general_complex_coefficient} 
\end{equation} or explicitly 
\begin{equation} C_{n,L} = \sum_{p,q\geq0} c_{pq} \sum_{\substack{ k_1+\cdots+k_p+l_1+\cdots+l_q=n\\ l_1+\cdots+l_q=L }} a_{k_1}\cdots a_{k_p} b_{l_1}\cdots b_{l_q}. 
\label{eq:general_CnL} 
\end{equation} 
The observable amplitude and phase are then 
\begin{equation} A^{\mathrm{obs}}_{n,L}=2|C_{n,L}|, \qquad \mathrm{and}  \qquad \phi^{\mathrm{obs}}_{n,L}=\arg(C_{n,L}). 
\label{eq:general_ampli_phase} 
\end{equation}
Equations~(\ref{eq:general_CnL})--(\ref{eq:general_ampli_phase}) show that, in the general non-linear case, the amplitude and phase of a given peak are not determined by a single parent component. Several different combinations of the harmonics of the two oscillations can fall at the same frequency $n\omega_0+L\Omega$, and their complex amplitudes add vectorially. Consequently, constructive or destructive interference between different non-linear paths can strongly affect the observed amplitudes and phases. 
The frequency positions alone do not distinguish between combined (AM+FM) modulation and general non-linear coupling, because both can produce peaks at $nf_0+Lf_{\mathrm{B}}$. The diagnostic information is therefore contained mainly in the amplitude and phase relations. 

\subsection{Several close frequencies} 

The formalism can be extended one step further by allowing more than one close secondary oscillation in the general non-linear case. Let the observed signal be an analytic function of a dominant oscillation and $J$ secondary oscillations, 
\begin{equation} X(t)=F[x_0(t),x_1(t),\ldots,x_J(t)] , 
\end{equation} where 
$\omega_j=\omega_0+\Omega_j$, $j=1,\ldots,J$. Expanding $F$ in a multivariate power series shows that every non-linear term generates integer linear combinations of the frequencies of the participating oscillations. Since each secondary frequency can be written as $\omega_j=\omega_0+\Omega_j$, any such combination can be expressed in the form 
\begin{equation} \omega_{n,\mathbf{L}} = n\omega_0+\sum_{j=1}^{J} L_j\Omega_j , 
\end{equation} 
or, in frequency units, 
\begin{equation} f_{n,\mathbf{L}} = n f_0+\sum_{j=1}^{J} L_j f_{\mathrm{B},j}, \qquad n,L_j\in\mathbb{Z},
\label{eq:multi_general_grid}
\end{equation} 
where only positive observed frequencies, $f_{n,\mathbf{L}}>0$, are retained in the Fourier spectrum of a
real-valued light curve.

This expression is the natural generalisation of Eq.~(\ref{eq:general_frequency_grid}) to several close frequencies. For $J>1$, however, the frequency pattern is no longer a one-dimensional multiplet sequence around each harmonic, but a multi-dimensional combination grid projected onto the observed frequency axis. If the beat frequencies $f_{\mathrm{B},j}$ are mutually incommensurable, the resulting side peaks need not form equidistant multiplets. If some of the beat frequencies are commensurable, different non-linear paths can fall at the same observed frequency, and their complex amplitudes must again be added vectorially. Equation~(\ref{eq:multi_general_grid}) should be interpreted again as the set of allowed frequencies only. The actual presence and strength of a given component
depend on the harmonic content of the parent oscillations and on the order and coefficients of the non-linear terms.
Thus, in the most general case, the frequency positions alone provide even less discriminatory power than in the single-secondary case. The useful diagnostics must again come from the amplitude hierarchy, the phase relations, and the identification of which non-linear paths can contribute to a given observed peak.
Resonance conditions of higher non-linear order correspond, in this language, to near coincidences between an intrinsic frequency and one of the allowed integer combinations in Eq.~(\ref{eq:multi_general_grid}).

\section{The diagnostic tool: \textsc{freqdiag}}

To ensure that the diagnostic methods described above and in B11 can be directly applied to the observed stars, we have summarised the main findings in Tables~\ref{tab:frequency-morphology} and \ref{tab:model-diagnostics}. Furthermore, we developed a simple, frequency-list-based classification program. The flowchart of the Python program \textsc{freqdiag}\footnote{The version of the \textsc{freqdiag} program used in this work is available at \url{https://github.com/benkoejozsef/Freqdiag/tree/v1.0.0-rc2}} is shown in Fig.~\ref{fig:flow}. The program relies on the NumPy and pandas libraries \citep{NumPy, Pandas}.

\subsection{Input data and decision scheme}

The program reads a list of significant Fourier components containing frequencies, amplitudes, and phases. All phases must refer to the same epoch. Phases obtained by consecutive pre-whitening can only be used if the fit is subsequently carried out for the entire frequency list; otherwise, incorrect results can be obtained.
After filtering and phase-convention conversion, it adopts an explicit frequency tolerance or derives it from the time base, $T$. The primary frequency, $f_0$, can be supplied by the user; otherwise it is selected by ranking candidate harmonic sequences. In the automatic ranking, one observed component can support at most one harmonic order. The program issues a warning when $f_0\leq1/T$, because adjacent harmonics are then not independently resolved.

The detected $nf_0$ harmonics are used to search for repeated close-frequency spacings around all available harmonic orders. Close spacings are clustered into $f_{\mathrm{B}}$ candidates, while integer-multiple relations are retained as candidate families rather than treated as independent proof of different physical frequencies. For each candidate, the program constructs the linear-secondary sequence $lf^{\prime}$, the low-frequency terms $lf_{\mathrm{B}}$, a neutral equidistant multiplet grid, the modulation grid $nf_0+Lf_{\mathrm{B}}$, and quadratic-coupling hypotheses anchored separately to the observed right- and left-side $f^{\prime}$ components.

The physical modulation--coupling comparison uses exactly the same
complex side peaks. For each signed side order $L$, the normalised complex observable is
\begin{equation}
    y_{n,L}=\frac{C(nf_0+Lf_{\mathrm B})}{C_n}.
\end{equation}
The modulation model has the form $y_{n,L}=a_L+n b_L$, whereas the quadratic-coupling model is $y_{n,L}=q_L x_{n,L}$, where $x_{n,L}$ is the corresponding shifted-primary--secondary product, with complex conjugation for a difference-frequency term. Thus, for each signed-$L$ group, the modulation and coupling models contain two and one complex free parameters, respectively, corresponding to four and two real parameters. A group is retained only if it contains at least three common side peaks, and model selection requires at least eight common complex data points in total. Each complex point contributes two real observations to the information criteria.

For either model, the normalised complex residual scatter is defined as
\begin{equation}
s =\left[\frac{\sum_i |y_i-\hat{y}_i|^2} {\sum_i |y_i|^2} \right]^{1/2},
\end{equation}
where $\hat{y}_i$ denotes the normalised complex value predicted by the fitted model.
We defined $\Delta\mathrm{AICc}=\mathrm{AICc}_{\mathrm{COUP}}-\mathrm{AICc}_{\mathrm{MOD}}$ and analogously for
$\Delta\mathrm{BIC}$. Here, AICc and BIC denote the corrected Akaike information criterion and the Bayesian information criterion, respectively. A directional preference is assigned only if both differences have the same sign and absolute values of at least 2; positive values favour modulation and negative values favour coupling. If the preferred model also has the smaller residual scatter, $s\leq0.35$ gives a \texttt{STRONG} classification, whereas $s>0.35$ gives \texttt{LEAN}. If the information-criterion winner has the larger scatter, the lower-scatter direction is retained but is capped at \texttt{LEAN}. Insufficient common data, differences below the adopted threshold, or disagreement between AICc and BIC result in \texttt{AMBIG}. The right- and left-side $f^{\prime}$ comparisons are then combined conservatively: agreement retains the weaker confidence, one directional anchor plus one ambiguous anchor yields a one-sided \texttt{LEAN} result, and a direct conflict remains \texttt{AMBIG}.
    
Two additional amplitude-order tests are applied hierarchically. Complete same-harmonic ($|L|=1,2,3$) tracks provide the higher-priority side-order test and can independently return \texttt{MOD-LEAN} or \texttt{COUP-LEAN} even if no common complex fit is available. If this test is non-directional, the code examines the individual signed $L=2$ and $L=3$ amplitude sequences along harmonic order $n$; this lower-priority test has only \texttt{LEAN} authority and requires a usable common fit to change the classification. \texttt{BL-MODLIKE} is used only when a dense multiplet structure rejects isolated linear beating, a simple stationary modulation fit remains poor, all usable coupling alternatives are disfavoured or absolutely poor, and neither side-order test selects coupling. With the default settings, \texttt{BL-MODLIKE} additionally requires at least four matched multiplet side peaks with a grid coverage of at least 0.5; a coupling fit is regarded as `absolutely poor' when $s>0.75$. Candidates with $|f_{\mathrm{B}}|<1/T$ are retained as frequency-grid matches but are labelled unresolved and cannot receive a headline classification.

The scope of the implemented coupling tests is narrower than that of the general coupling formalism developed in Sects.~\ref{sec:nonlin_coupl} and \ref{Sec:general}. The fitted \texttt{COUP} hypotheses include only the primary--secondary quadratic-coupling relations of Sects.~\ref{Sec:quadratic_sin}--\ref{Sec:quad_multi_independent}. The secondary--secondary terms discussed in Sect.~\ref{Sec:quadratic_multi_general} are checked only at the level of their allowed frequency positions and do not enter the amplitude--phase model comparison or physical ranking. Although amplitudes and phases could in principle be used to test a specified finite-order coupling model, the unrestricted expansion of Sect.~\ref{Sec:general} contains multiple non-linear paths and unspecified coupling coefficients contributing to the same Fourier component. It therefore yields no universal amplitude-ratio, phase, or side-order relation that could be implemented as a model-independent classifier. Consequently, \texttt{COUP} denotes compatibility only with the implemented primary--secondary quadratic model, whereas \texttt{AMBIG} neither supports nor excludes more general non-linear coupling.

\subsection{Validation on test frequency lists}

We tested the program on frequency lists derived from synthetic and observed light curves with different time bases and, consequently, different Fourier resolutions. From the Fourier parameters of the artificial light curves shown in this article ($T=200$~d), we were able to reproduce the actual input beating and coupling cases in every instance. We then considered observed stars, using both our own frequency tables generated with different frequency-analysis programs and published frequency solutions, demonstrating the applicability of \textsc{freqdiag} to inputs obtained with different tools. Table~\ref{tab:freqdiag_observed_samples} shows the details of the test results for the observed data\footnote{The frequency lists used in this study are available in the \texttt{test} directory of the GitHub repository.}. 

We applied \textsc{freqdiag} to the frequency lists of 15 \textit{Kepler} Blazhko RRab stars ($T=1440$ or $650$~d), derived from the tailor-made light curves published by \citet{Benko2014}.
The diagnostic tool favours modulation profiles for 13 stars. For two stars exhibiting highly complex light variations (V2178 Cyg and V445 Lyr), our results suggest that modulation-like variations and non-linear coupling are present simultaneously. In the case of V445 Lyr, for example, the Fourier characteristics of the primary Blazhko signal are modulation-like, while the secondary Blazhko frequency can be explained by coupling. These mixed cases require further investigation. 

For the six Blazhko RRc stars in the TESS continuous viewing zone (CVZ) studied by \citet{Benko2023} ($T=350$~d), the program returned classifications in four cases, all favouring modulation.
We also tested the program on ten Blazhko RRab and ten Blazhko RRc stars observed by K2 ($T=90$~d). The program returned modulation-like classifications for nine RRab stars; for the tenth, it retained only a neutral multiplet-grid match. Among the RRc stars, only one received a modulation-like classification, whereas the remaining nine were ambiguous. This result illustrates that nearly sinusoidal light curves, as are common among RRc stars, 
are less informative for the diagnosis because they provide only a few detectable harmonics.

We also tested published frequency solutions for two HADS stars.
Although the frequency solution of CoRoT 101155310 \citep[$T=152$~d;][]{Poretti2011} contains ten harmonics of the dominant pulsation, only four corresponding side peaks are detected, around $f_0$ and $2f_0$. These are insufficient for a constrained common modulation--coupling fit, and the result is therefore \texttt{AMBIG}. For the double-mode HADS star KIC~10284901 \citep[$T=284$~d;][]{Yang_Esamdin2019}, we analysed the published 151-frequency solution in \texttt{frequency-only} mode because its table does not include phases. \textsc{freqdiag} recovered the two reported `modulation' frequencies with complete first-order side-peak grids around $f_0$ and $2f_0$. Both modulation candidates nevertheless remained \texttt{AMBIG}, because neither phase information nor sufficiently extended higher-side-order sequences are available.

These examples show that a rich harmonic spectrum alone does not guarantee a decisive diagnosis: sufficiently many side peaks, consistently fitted phases, and coverage across several harmonic and side orders are also required. The long, homogeneous time series expected from PLATO \citep{Rauer2025} for a substantially larger sample of HADS stars should allow more complete frequency solutions to be derived directly from the light curves and may therefore permit more decisive applications of the method.

We generated the \textit{Kepler} and TESS frequency lists using our Python code, which performs block-based pre-whitening and computes Lomb--Scargle amplitude spectra. The program was originally developed for the analysis of the \textit{Kepler} Blazhko sample. The main aim was to minimise the number of pre-whitening steps; therefore, it searched for harmonics, side peaks, etc. within separate blocks, and then pre-whitened each block in a single step. We applied an $S/N\geq4$ criterion to the primary frequency, its harmonics, and the primary Blazhko and combination components and an $S/N\geq5$ criterion for further independent frequencies and a possible second, closely spaced family of frequencies. Following the frequency selection, all retained components were simultaneously fitted to the original light curve using the linear least squares method. The phases in the output frequency table are given in cycles under the cosine convention. 

The K2 frequency lists were obtained with the MultiFrequencyFitter module of \textsc{seismolab} \citep{Bodi2024} using iterative Lomb–Scargle pre-whitening. All retained frequencies, amplitudes, and phases were simultaneously refitted to the full light curve. The phases were expressed in radians using the sine convention.

As a complementary validation, we analysed 33 synthetic light curves with known AM, FM, or combined modulation. These artificial light curves are those presented in the figures of B11. Their frequency lists were obtained in the same way as those of the \textit{Kepler} stars. To test the physical classification independently of the candidate-search stage, we supplied the known modulation frequency  directly and evaluated both coupling anchors. Of the 33 inputs, 25 were classified as \texttt{MOD-STRONG}, four as \texttt{MOD-LEAN}, and one as \texttt{BL-MODLIKE}. The remaining three were \texttt{AMBIG} because their extracted frequency lists contained insufficient common side-peak information for model comparison. None was classified as beating or coupling.

\subsection{Dependence on time base and cycle coverage}

\begin{figure}
    \centering
    \includegraphics[width=\linewidth]{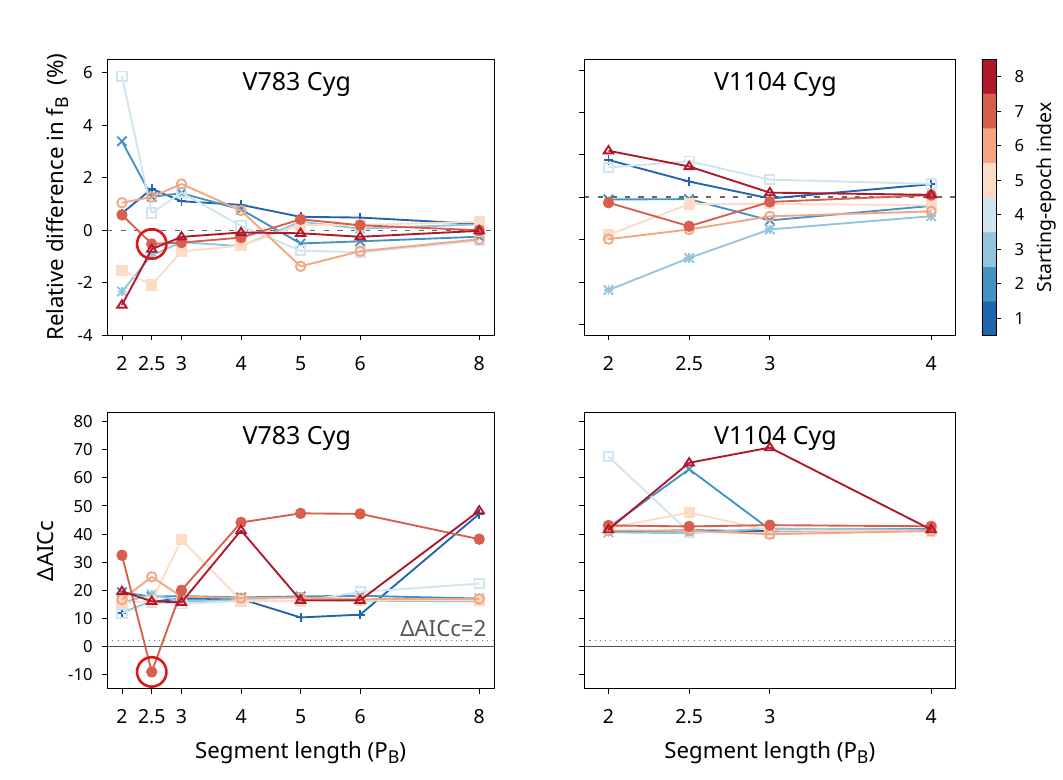}
    \caption{Dependence of the \textsc{freqdiag} results on segment length and starting epoch for V783 Cyg (\textit{left}) and V1104 Cyg (\textit{right}). The colours and symbols identify the eight starting epochs. \textit{Top}: Relative difference between the close-frequency spacing recovered from each segment and that obtained from the full \textit{Kepler} time series. \textit{Bottom}: More conservative of the AICc differences obtained with the right- and left-side coupling anchors, where positive values favour modulation. The single \texttt{AMBIG} result, obtained for a 2.5-cycle segment of V783 Cyg, is highlighted with a red circle.}
    \label{fig:segment_test}
\end{figure}
To examine how the time base and the starting epoch affect the result, we divided the well-sampled \textit{Kepler} light curve of V783 Cyg into segments spanning 2, 2.5, 3, 4, 5, 6, and 8 Blazhko cycles, and that of V1104 Cyg into segments spanning 2, 2.5, 3, and 4 cycles. For each segment length, we used eight starting epochs and repeated the complete frequency determination before applying \textsc{freqdiag}. All 32 segments of V1104 Cyg and 55 of the 56 segments of V783 Cyg were classified as \texttt{MOD-STRONG}. Only one 2.5-cycle segment of V783 Cyg produced an \texttt{AMBIG} result because the two coupling-anchor comparisons were inconsistent and only a few common complex side peaks were available (see Fig.~\ref{fig:segment_test}). Thus, in these harmonically rich RRab spectra, a stable result can usually be obtained from approximately two cycles, although an individual starting epoch can still yield an ambiguous solution. This value is therefore an empirical lower limit for the RRab stars tested here rather than a generally applicable threshold.

By contrast, the harmonically sparse RRc samples can remain under-constrained even when 9--12 cycles are covered. The diagnostic information is therefore determined jointly by $T f_{\mathrm{B}}$, the number of detected primary harmonics, the available signed side orders, and the number of common complex side peaks. Generally speaking, the simpler the spectrum, the longer the data series required to make a well-founded decision.

\section{Conclusions}

The Fourier spectra produced by modulation, linear beating, and non-linear coupling form a nested hierarchy rather than three disjoint pattern classes. Strictly linear beating is the most restrictive case: a sinusoidal close secondary produces a single additional peak, while a non-sinusoidal secondary produces its own harmonic sequence ($lf'=lf_0+lf_{\mathrm{B}}$), whose separation from the primary harmonics increases with harmonic order. Quadratic coupling generates combination terms ($nf_0+lf_{\mathrm{B}}$), with amplitudes and phases tied to shifted primary harmonics and to the parent secondary oscillation. At arbitrary non-linear order, the allowed frequencies occupy the same grid as a modulated signal, so frequency positions alone cannot establish the physical origin of a multiplet.

The useful information is therefore carried by relations among complex amplitudes and phases and by their behaviour across harmonic and side orders. Our \textsc{freqdiag} program combines these relations conservatively: modulation and coupling are fitted to the same normalised complex side peaks, right- and left-side $f^{\prime}$ coupling anchors are combined by consensus, and two hierarchical side-order tests can modify the preferred classification. The coupling classification has a deliberately restricted scope. A \texttt{COUP} label denotes preference for the implemented primary--secondary quadratic-coupling model; secondary--secondary terms are tested only at the frequency-grid level, and general non-linear coupling is not classified. Thus, \texttt{COUP} should not be interpreted as a classification of non-linear coupling in general, and an \texttt{AMBIG} result does not rule out such coupling.

In the test samples, modulation-like solutions dominate the RRab stars: they are obtained for 13 of 15 \textit{Kepler} targets and 9 of 10 K2 targets. Two \textit{Kepler} RRab stars exhibit mixed characteristics; one K2 RRab target retains only a neutral multiplet-grid classification. The RRc results are less decisive: four of six TESS CVZ but only one of ten K2 stars receive a modulation-like classification, with the remaining cases under-constrained. This contrast is consistent with the much poorer harmonic and common-side-peak support of nearly sinusoidal RRc light curves.

We find that no universal minimum number of secondary cycles guarantees a successful diagnosis. Nevertheless, long observational baselines are essential, not only because they cover more cycles, but also because they resolve narrow frequency spacings and recover the harmonics and side peaks required by the diagnostic tests.

These results demonstrate that modulation, linear beating, and non-linear coupling are not necessarily observationally indistinguishable. When a Fourier spectrum contains sufficiently many harmonics and side peaks, their different amplitude--phase relations and side-order behaviour provide a practical route to separating the competing models. Although the output classes should be interpreted as indicating degrees of model compatibility rather than as definitive physical identifications, the resulting classifications can identify stars whose Blazhko-like variability is more naturally associated with one of the different mechanisms. Applied to larger samples, this approach may help to establish whether the phenomenological Blazhko class is physically homogeneous or not.

\begin{acknowledgements}
The research was partially supported by the `SeismoLab' KKP-137523 \'Elvonal grant of the Hungarian Research, Development and Innovation Office (NKFIH). OpenAI ChatGPT Plus 5.6Sol was used to assist in developing the Python code for the synthetic datasets and the \textsc{freqdiag} tool. All generated code and outputs were reviewed and validated by the authors.

\end{acknowledgements}

\bibliographystyle{aa}
\bibliography{Beating}

\FloatBarrier
\begin{appendix}

\onecolumn
\newcommand{\freqdiaggrouphead}[1]{
  \addlinespace[2pt]
  \multicolumn{4}{@{}l}{\textsc{#1}}\\[-1pt]
  \midrule
}

\section{Summary of observational diagnostics}\label{app:diagnostics}

Tables~\ref{tab:frequency-morphology} and \ref{tab:model-diagnostics} summarise the main observational differences between modulation, beating, and non-linear coupling.

\begin{table}[!ht]
\caption{Frequency-domain morphology expected for modulation, linear beating, and non-linear coupling.}
\label{tab:frequency-morphology}
\centering
\small
\setlength{\tabcolsep}{3.2pt}
\renewcommand{\arraystretch}{1.13}
\begin{tabularx}{\textwidth}{@{}>{\RaggedRight\arraybackslash}p{0.19\textwidth}*{3}{>{\RaggedRight\arraybackslash}X}@{}}
\toprule
Model & Main Fourier pattern & Low-frequency components & Separation pattern \\
\midrule
\freqdiaggrouphead{Modulation}
Sinusoidal AM
  & $n f_0\pm f_{\mathrm{m}}$; only first-order side peaks (triplets).
  & $f_{\mathrm{m}}$ can occur if the mean level is also modulated.
  & $\Delta f=f_{\mathrm{m}}$ at every harmonic. \\
Non-sinusoidal AM
  & $n f_0\pm r f_{\mathrm{m}}$ for the detectable harmonics $r=1,\ldots,R_{\mathrm{A}}$ of the AM waveform.
  & Terms $r f_{\mathrm{m}}$ can occur if the mean level is also modulated.
  & $\Delta f=r f_{\mathrm{m}}$; a finite set of equidistant side orders. \\
Sinusoidal FM/PM
  & In principle, an infinite multiplet $n f_0+L f_{\mathrm{m}}$, $L=0,\pm1,\pm2,\ldots$, around every harmonic.
  & No
  & $\Delta f=|L|f_{\mathrm{m}}$, independent of $n$. \\
Non-sinusoidal FM/PM
  & $n f_0+L f_{\mathrm{m}}$, $L=0,\pm1,\pm2,\ldots$; even a finite Fourier representation of the phase waveform generally produces infinitely many side orders.
  & No
  & The same equidistant grid; several harmonics of the phase waveform contribute to a given $L$. \\
Sinusoidal AM+FM/PM
  & $n f_0+L f_{\mathrm{m}}$, $L=0,\pm1,\pm2,\ldots$; first-order AM combines with the FM/PM multiplet.
  & $f_{\mathrm{m}}$ can occur if the mean level is amplitude-modulated.
  & Constant spacing $|L|f_{\mathrm{m}}$ for all harmonic orders. \\
Non-sinusoidal AM+FM/PM
  & $n f_0+L f_{\mathrm{m}}$, $L=0,\pm1,\pm2,\ldots$; the AM and FM/PM Fourier series are convolved.
  & Terms $r f_{\mathrm{m}}$ can occur through mean-level AM.
  & One equidistant grid, with side orders populated by several AM and FM/PM contributions. \\
Multiple modulation frequencies
  & In general, $n f_0+\sum_j L_j f_{\mathrm{m},j}$; pure additive AM is the restricted case $n f_0\pm r f_{\mathrm{m},j}$.
  & Individual $r f_{\mathrm{m},j}$ terms can occur through mean-level AM; mixed low-frequency combinations require cross terms.
  & Several offset families; commensurate offsets can coincide, while incommensurate offsets form a projected multi-dimensional grid. \\

\freqdiaggrouphead{Linear beating}
One sinusoidal secondary
  & Primary harmonics $k f_0$ plus one close frequency $f'=f_0+f_{\mathrm{B}}$.
  & No $f_{\mathrm{B}}$ term is generated by linear superposition.
  & One close peak near $f_0$; no corresponding peaks around higher harmonics. \\
One non-sinusoidal secondary
  & Two independent sequences: $k f_0$ and $l f'=l f_0+l f_{\mathrm{B}}$.
  & No $f_{\mathrm{B}}$ term
  & Separation from the corresponding primary harmonic, $lf_0$, grows as $|l f_{\mathrm{B}}|$. \\
Several secondaries
  & $k f_0$ plus sequences $l f'_j=l f_0+l f_{\mathrm{B},j}$.
  & Neither $f_{\mathrm{B},j}$ nor $|f'_i-f'_j|$ is generated.
  & Several $|l f_{\mathrm{B},j}|$ spacings; the apparent side peaks need not be equidistant. \\

\freqdiaggrouphead{non-linear coupling}
Quadratic; one sinusoidal secondary
  & Combination terms $n f_0\pm f_{\mathrm{B}}$, including the difference frequency $f_{\mathrm{B}}$.
  & $f_{\mathrm{B}}$ is generated.
  & Constant first-order spacing, but amplitudes are tied to neighbouring primary harmonics. \\
Quadratic; one non-sinusoidal secondary
  & $n f_0+l f_{\mathrm{B}}$ for each harmonic $l$ present in the secondary signal.
  & Terms $l f_{\mathrm{B}}$ can be generated.
  & Constant spacing for a fixed $l$; the maximum side order is limited by the secondary harmonic content. \\
Several secondaries coupled only to the primary
  & One system $n f_0+l f_{\mathrm{B},j}$ for each secondary.
  & Terms $l f_{\mathrm{B},j}$ can occur, but no secondary--secondary difference terms.
  & Multiple, generally non-equidistant systems. \\
General quadratic coupling
  & Primary--secondary terms plus cross-combinations such as $2f_0+f_{\mathrm{B},i}+f_{\mathrm{B},j}$.
  & Differences such as $|f_{\mathrm{B},i}-f_{\mathrm{B},j}|$ can occur.
  & Several overlapping frequency families; no unique separation pattern. \\
General non-linear coupling
  & $n f_0+L f_{\mathrm{B}}$ for one secondary, or $n f_0+\sum_j L_j f_{\mathrm{B},j}$ for several.
  & Any low-frequency combination allowed by the non-linear order can occur.
  & The grid can be identical to a modulation grid; positions alone are not diagnostic. \\
\bottomrule
\end{tabularx}
\tablefoot{Here $f_0$ is the dominant pulsation frequency, $f'$ is a close secondary frequency, $f_{\mathrm{B}}=f'-f_0$, and $f_{\mathrm{m}}$ is the modulation frequency. In the modulation rows, $R_{\mathrm{A}}$ is the highest detectable harmonic of the AM waveform. The term `side peak' is descriptive; in a beating model it does not imply modulation.}
\end{table}

\begin{table}[p]
\caption{Amplitude--phase diagnostics and principal limitations of the models in Table~\ref{tab:frequency-morphology}.}
\label{tab:model-diagnostics}
\centering
\small
\setlength{\tabcolsep}{3.2pt}
\renewcommand{\arraystretch}{1.1}
\begin{tabularx}{\textwidth}{@{}>{\RaggedRight\arraybackslash}p{0.19\textwidth}*{3}{>{\RaggedRight\arraybackslash}X}@{}}
\toprule
Model & Amplitude and phase relation & Most useful discriminator & Main caveat \\
\midrule
\freqdiaggrouphead{Modulation}
Sinusoidal AM
  & The two first-order side peaks are equal in amplitude and scale with the parent amplitude $A_n$; their phases are fixed by the parent and AM phases.
  & Symmetric triplets with a common spacing and $A_n$ scaling.
  & Asymmetry requires an additional effect, such as FM/PM or harmonic-dependent modulation. \\
Non-sinusoidal AM
  & For every AM harmonic $r$, the $\pm r$ pair is symmetric and $A_{n,\pm r}\propto A_n M_r$; the phase offsets follow the phase of that AM harmonic.
  & A finite set of symmetric side-order pairs, each reproducing the parent-harmonic amplitude progression.
  & The finite-order prediction assumes that the AM waveform itself has no detectable harmonics above $R_{\mathrm{A}}$. \\
Sinusoidal FM/PM
  & For a common sinusoidal phase perturbation, $A_{n,L}\propto A_n|J_L(n\beta)|$; the phases follow the corresponding Bessel and modulation-phase relations.
  & The Bessel hierarchy of symmetric $\pm L$ pairs across both harmonic order ($n$) and side order ($L$).
  & Bessel amplitudes need not decrease monotonically and can pass through zeros. \\
Non-sinusoidal FM/PM
  & The complex coefficient of side order $L$ is a generalised-Bessel sum over all harmonic combinations of the phase waveform that yield $L$.
  & One periodic phase-modulation waveform must reproduce the complex side peaks around all primary harmonics.
  & The $+L$ and $-L$ amplitudes need not be equal, and the side-order envelope need not be monotonic. \\
Sinusoidal AM+FM/PM
  & AM and FM/PM contributions add as complex coefficients; their relative strength and phase can produce strongly asymmetric multiplets.
  & A common complex AM+FM/PM solution across harmonics, including the relative AM--FM phase.
  & Frequency positions alone are degenerate with general non-linear coupling. \\
Non-sinusoidal AM+FM/PM
  & The side-peak coefficients are convolutions of the AM Fourier coefficients and generalised FM/PM coefficients.
  & A single pair of periodic AM and phase waveforms must reproduce all harmonic and side-order sequences simultaneously.
  & The model is flexible: asymmetric and non-monotonic side orders can resemble non-linear coupling. \\
Multiple modulation frequencies
  & Each peak coefficient belongs to a multi-dimensional modulation lattice and depends on the amplitudes and phases of all contributing modulation components.
  & The same offset families and their complex-amplitude relations recur around many primary harmonics and agree with the recovered modulation periods.
  & Dense projections, commensurabilities, and limited frequency resolution can mimic several coupled secondary oscillations. \\

\freqdiaggrouphead{Linear beating}
One sinusoidal secondary
  & The fundamental shows apparent amplitude and phase variation, while the primary harmonics remain unchanged.
  & Apparent time-domain AM/PM without a repeated Fourier multiplet.
  & A very close or unresolved secondary can be difficult to separate from a trend. \\
One non-sinusoidal secondary
  & Components have the independent amplitudes and phases of the two parent harmonic sequences.
  & Separation increases linearly with harmonic order.
  & Coincident or unresolved components add vectorially. \\
Several secondaries
  & Each secondary has its own harmonic sequence; coincident components add as complex amplitudes.
  & Non-equidistant close peaks without difference or cross-combination frequencies.
  & Missing combination peaks constrain coupling only if they should be detectable above the local noise. \\

\freqdiaggrouphead{non-linear coupling}
Quadratic; one sinusoidal secondary
  & $A^+_n\propto A_{n-1}B$ and $A^-_n\propto A_{n+1}B$; phases obey combination relations.
  & Asymmetric triplets tied to neighbouring, rather than parent, harmonic amplitudes.
  & Intrinsic secondary peaks and combination peaks require consistent anchor treatment. \\
Quadratic; one non-sinusoidal secondary
  & Away from intrinsic secondary harmonics, $A^{\mathrm{obs}}_{n,l}\propto A_{n-l}B_l$ and $\phi^{\mathrm{obs}}_{n,l}=\phi_{n-l}+\psi_l+\arg\alpha$.
  & Shifted-primary amplitude tracks and approximately constant combination phases.
  & RRab harmonic-amplitude reversals can produce non-monotonic side-order tracks; the harmonic range must extend beyond the reversal. \\
Several secondaries coupled only to the primary
  & Each family obeys the single-secondary combination relations; coincident terms add vectorially.
  & Separate multiplet families with separate combination-phase relations.
  & Commensurate offsets can merge several non-linear paths into one observed peak. \\
General quadratic coupling
  & Cross-coupling can enhance or suppress peaks through constructive or destructive interference.
  & A low-frequency secondary--secondary difference together with the corresponding cross-combination peaks.
  & A low-frequency peak by itself does not identify the non-linear path. \\
General non-linear coupling
  & Every peak is the vector sum of all non-linear paths reaching the same frequency.
  & Amplitude hierarchy, phase relations, and identifiable parent combinations used jointly.
  & The inverse problem is non-unique; classifications express graded model compatibility. \\
\bottomrule
\end{tabularx}
\tablefoot{In the modulation rows, $M_r$ is the amplitude of the $r$th harmonic of the AM waveform, $R_{\mathrm{A}}$ is its highest detectable harmonic, and $\beta$ is the sinusoidal phase-modulation index. The general multiple-modulation expression also covers combination side peaks generated by simultaneous modulation frequencies.}
\end{table}

\begin{figure}
    \centering
    \includegraphics[width=\linewidth]{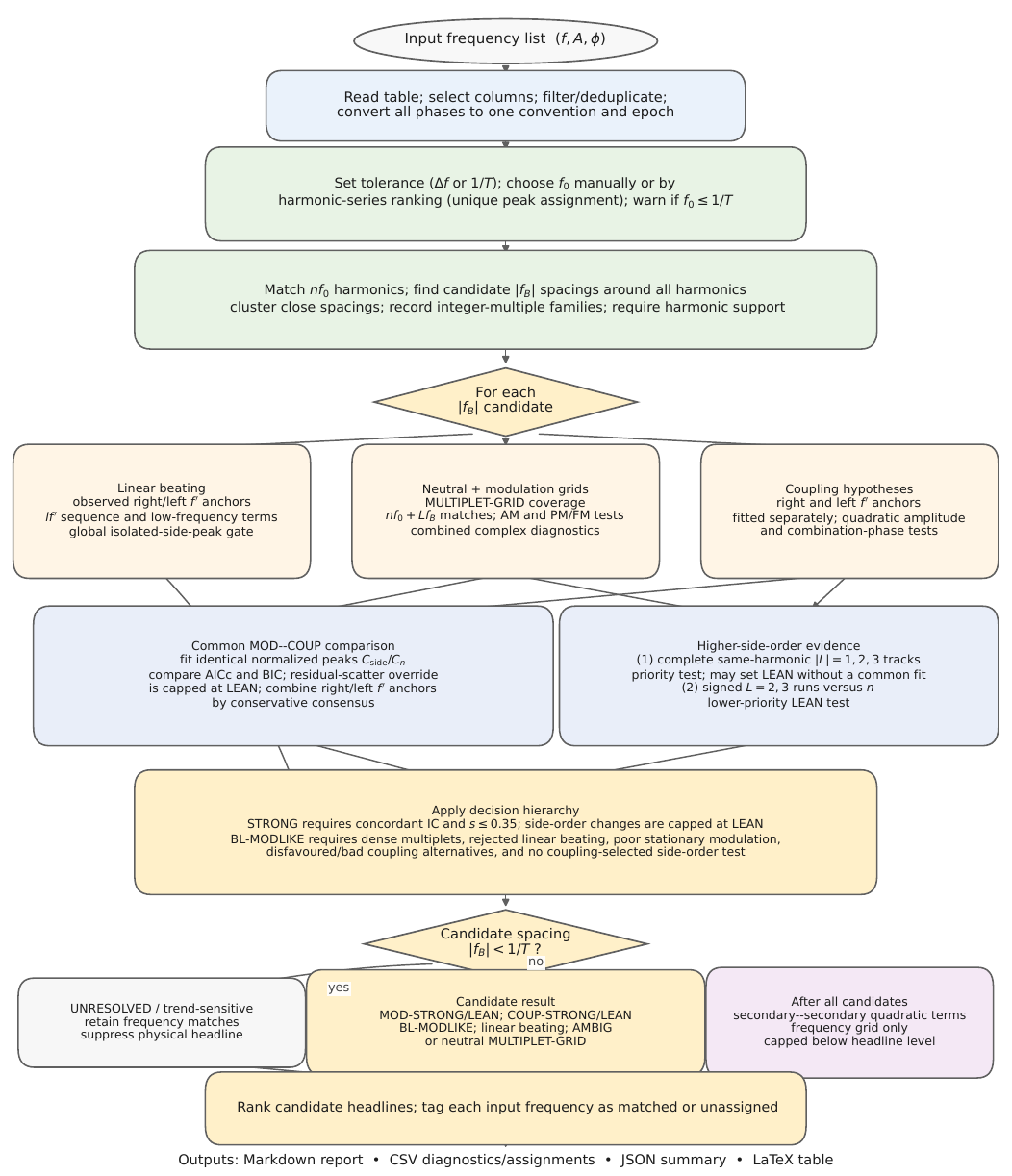}
    \caption{Flowchart of \textsc{freqdiag}, which evaluates modulation, beating, and coupling hypotheses from Fourier frequency lists at the adopted frequency resolution.
    }
    \label{fig:flow}
\end{figure}

\FloatBarrier

\begin{table}[p]
\caption{\textsc{freqdiag} results for the observed test samples.}
\label{tab:freqdiag_observed_samples}
\centering
\setlength{\tabcolsep}{4pt}
\begin{tabular}{@{}lrrrrl@{}}
\hline\hline
\noalign{\smallskip}
Star or ID
  & $T$ (d)
  & $P_0$ (d)
  & $P_{\mathrm{c}}$ (d)
  & $N_{\mathrm{c}}$
  & Result \\
\noalign{\smallskip}
\hline
\noalign{\smallskip}
\multicolumn{6}{@{}l}{\textit{Kepler} RRab stars} \\
V2178 Cyg 
  & 1440 & 0.486947 & 85.25 & 16.89 & \texttt{BL-MODLIKE} \\
  &      &          & 220.79 & 6.52 & \texttt{COUP-LEAN} \\
V808 Cyg 
  & 1440 & 0.547863 & 92.03 & 15.65 & \texttt{MOD-LEAN} \\
V783 Cyg 
  & 1440 & 0.620700 & 27.67 & 52.05 & \texttt{MOD-LEAN} \\
V354 Lyr 
  & 1440 & 0.561689 & 34.28 & 42.00 & \texttt{MOD-STRONG} \\
V445 Lyr 
  & 1440 & 0.513090 & 145.36 & 9.91 & \texttt{COUP-LEAN} \\
  &      &          & 55.03 & 26.17 & \texttt{MOD-LEAN} \\
KIC 7257008
  & 650 & 0.511787 & 39.47 & 16.47 & \texttt{MOD-LEAN} \\
V355 Lyr 
  & 1440 & 0.473700 & 31.06 & 46.36 & \texttt{MOD-STRONG} \\
V450 Lyr 
  & 1440 & 0.504619 & 121.07 & 11.89 & \texttt{MOD-STRONG} \\
V353 Lyr 
  & 1440 & 0.556800 & 71.69 & 20.09 & \texttt{MOD-LEAN} \\
V366 Lyr 
  & 1440 & 0.527028 & 62.88 & 22.90 & \texttt{BL-MODLIKE} \\
V360 Lyr 
  & 1440 & 0.557576 & 52.09 & 27.64 & \texttt{MOD-STRONG} \\
KIC 9973633
  & 650 & 0.510783 & 67.25 & 9.67 & \texttt{MOD-STRONG} \\
V838 Cyg 
  & 1440 & 0.480280 & 59.41 & 24.24 & \texttt{BL-MODLIKE} \\
KIC 11125706
  & 1440 & 0.613220 & 40.22 & 35.80 & \texttt{MOD-LEAN} \\
V1104 Cyg 
  & 1440 & 0.436385 & 52.01 & 27.69 & \texttt{MOD-LEAN} \\
\noalign{\smallskip}
\hline
\noalign{\smallskip}
\multicolumn{6}{@{}l}{TESS CVZ RRc stars} \\
Gaia DR2 4654252618960823936
  & 350 & 0.332472 & 48.63 & 7.20 & \texttt{MOD-LEAN} \\
Gaia DR2 4676321123000662272
  & 350 & 0.300572 & 8.45 & 41.40 & \texttt{MOD-STRONG} \\
NSVS 2852763
  & 350 & 0.315066 & 57.98 & 6.04 & \texttt{MOD-STRONG} \\
TYC 8896-623-1
  & 350 & 0.374491 & 88.89 & 3.94 & \texttt{AMBIG} \\
V420 Dra
  & 350 & 0.329545 & 118.47 & 2.95 & \texttt{MOD-STRONG} \\
XX Dor
  & 350 & 0.328935 & 13.60 & 25.74 & \texttt{AMBIG} \\
\noalign{\smallskip}
\hline
\noalign{\smallskip}
\multicolumn{6}{@{}l}{K2 RRab stars} \\
EPIC 201630427
  & 90 & 0.500909 & 35.91 & 2.51 & \texttt{MOD-STRONG} \\
EPIC 205905693
  & 90 & 0.486380 & 24.98 & 3.60 & \texttt{MOD-STRONG} \\
EPIC 210746706
  & 90 & 0.483327 & 32.87 & 2.74 & \texttt{MOD-LEAN} \\
EPIC 211518789
  & 90 & 0.504372 & 33.38 & 2.70 & \texttt{MOD-LEAN} \\
EPIC 212133146
  & 90 & 0.485634 & 29.74 & 3.03 & \texttt{MOD-STRONG} \\
EPIC 212749007
  & 90 & 0.531009 & 21.77 & 4.13 & \texttt{MOD-STRONG} \\
EPIC 212789806
  & 90 & 0.496870 & 163.91 & 0.55
  & \texttt{MULTIPLET-GRID}\tablefootmark{a} \\
EPIC 213624747
  & 90 & 0.548088 & 36.31 & 2.48 & \texttt{MOD-STRONG} \\
EPIC 249799246
  & 90 & 0.535499 & 30.55 & 2.95 & \texttt{MOD-STRONG} \\
EPIC 249923525
  & 90 & 0.653042 & 39.36 & 2.29 & \texttt{MOD-STRONG} \\
\noalign{\smallskip}
\hline
\noalign{\smallskip}
\multicolumn{6}{@{}l}{K2 RRc stars} \\
EPIC 204140546
  & 90 & 0.310306 & 7.37 & 12.21 & \texttt{AMBIG} \\
EPIC 220497567
  & 90 & 0.302151 & 9.26 & 9.72 & \texttt{AMBIG} \\
EPIC 246233565
  & 90 & 0.295455 & 7.48 & 12.04 & \texttt{AMBIG} \\
EPIC 248449270
  & 90 & 0.291016 & 22.38 & 4.02 & \texttt{AMBIG} \\
EPIC 249436580
  & 90 & 0.342800 & 39.09 & 2.30 & \texttt{MOD-LEAN} \\
EPIC 249789797
  & 90 & 0.289895 & 31.20 & 2.88 & \texttt{AMBIG} \\
EPIC 249883675
  & 90 & 0.324353 & 49.53 & 1.82 & \texttt{AMBIG} \\
EPIC 250003296
  & 90 & 0.354977 & 17.87 & 5.04 & \texttt{AMBIG} \\
EPIC 251529654
  & 90 & 0.266484 & 12.47 & 7.21 & \texttt{AMBIG} \\
EPIC 251809798
  & 90 & 0.259774 & 17.16 & 5.24 & \texttt{AMBIG} \\
\noalign{\smallskip}
\hline
\noalign{\smallskip}
\multicolumn{6}{@{}l}{ HADS stars} \\
CoRoT 101155310
  & 152 & 0.125799 & 5.18 & 29.37 & \texttt{AMBIG} \\
KIC 10284901
  & 284 & 0.052648 & 2.27 & 125.09 & \texttt{AMBIG} \\
  &     &          & 1.23 & 230.63 & \texttt{AMBIG} \\    
\noalign{\smallskip}
\hline
\end{tabular}
\tablefoot{$T$ is the approximate length of the observed time series used in classification.
$P_0=1/f_0$ is the adopted pulsation period. The characteristic period $P_{\mathrm{c}}=1/|\Delta f|$ is calculated from the close-frequency spacing tested by \textsc{freqdiag}; its notation is deliberately neutral. The number of covered characteristic cycles is $N_{\mathrm{c}}=T|\Delta f|$. For the two mixed \textit{Kepler} cases, V2178 Cyg and V445 Lyr, the two successive rows give the characteristic periods corresponding to the two different solutions.
\tablefoottext{a}{The candidate spacing is below the Fourier resolution, $|\Delta f|<1/T$; only a neutral frequency-grid match is retained.}
Here, \texttt{COUP} refers only to the primary--secondary quadratic coupling model implemented in \textsc{freqdiag}; \texttt{AMBIG} does not exclude more general non-linear coupling.}
\end{table}

\end{appendix}

\end{document}